\documentclass[fleqn,usenatbib]{mnras}

\usepackage{mathptmx,journals}

\usepackage[T1]{fontenc}
\usepackage{ae,aecompl}

\usepackage{graphicx}	
\usepackage{amsmath}	
\usepackage{amssymb}	

\title[IR interferometry of X-ray Binaries]{Infrared interferometry to spatially and spectrally resolve jets in X-ray binaries}

\author[Markoff et al. ]{Sera Markoff$^{1}$\thanks{E-mail:
s.b.markoff@uva.nl}, David M. Russell$^{2}$, Jason Dexter$^{3,4}$, Oliver Pfuhl$^{5}$,
\newauthor
Frank Eisenhauer$^{4}$, Roberto Abuter$^{5}$, James C.A. Miller-Jones$^6$, Thomas D. Russell$^{1}$\\
$^{1}$Anton Pannekoek Institute for Astronomy \& GRAPPA, University of
  Amsterdam, Science Park 904, 1098 XH Amsterdam, the Netherlands\\
$^{2}$Center for Astro, Particle and Planetary Physics, New York University Abu Dhabi, PO Box 129188, Abu Dhabi, UAE\\
$^{3}$JILA and Department of Astrophysical and Planetary Sciences, University of Colorado, Boulder, CO 80309, USA\\
$^{4}$Max Planck Institute for Extraterrestrial Physics (MPE), Giessenbachstr. 1, 85748 Garching, Germany\\
$^{5}$European Southern Observatory, Karl-Schwarzschild-Str. 2, 85748 Garching, Germany\\
$^{6}$International Centre for Radio Astronomy Research -- Curtin University, GPO Box U1987, Perth, WA 6845, Australia
}

\date{Accepted XXX. Received YYY; in original form ZZZ}

\pubyear{2019}

\begin{document}
\label{firstpage}
\pagerange{\pageref{firstpage}--\pageref{lastpage}}
\maketitle

\begin{abstract}
  Infrared interferometry is a new frontier for precision ground
  based observing, with new instrumentation achieving milliarcsecond (mas)
  spatial resolutions for faint sources, along with astrometry on the
  order of 10 microarcseconds ($\mu$as).  
  This technique has already led to breakthroughs in the observations of the supermassive black hole at the Galactic centre and its orbiting stars, AGN, and exo-planets, and can be employed for studying X-ray binaries (XRBs), microquasars in particular.  Beyond 
  constraining the orbital parameters of the system using the centroid
  wobble and spatially resolving jet discrete ejections on mas scales, 
  we also propose a novel method to discern between the
  various components contributing to the infrared bands:  accretion
  disk, jets and companion star.   
  We demonstrate that the GRAVITY instrument on the Very Large
  Telescope Interferometer (VLTI) should be able to detect a centroid shift
  in a number of sources, opening a new avenue of exploration for the myriad
  of transients expected to be discovered in the coming decade of
  radio all-sky surveys. We also present the first proof-of-concept
  GRAVITY observation of a low-mass X-ray binary transient, MAXI J1820+070, to search 
  for extended jets on mas scales.  We place the tightest  
  constraints yet via direct imaging on the size of the infrared emitting region of the compact jet in a hard state XRB.
\end{abstract}




\section{Introduction}

Radio wave interferometry has been in development for decades,
culminating in the exquisite precision of Very Long Baseline
Interferometry (VLBI).  However as one goes to higher frequencies,
atmospheric effects makes visibility corrections more challenging,
requiring generally shorter integration times on any given source.  In the optical/infrared (OIR) bands the previous generation of instruments could only image very bright sources using interferometry
 (e.g., V and H-band photometric
magnitudes of $\leq 2$; \citealt{Monnieretal2007,Cheetal2011}). Current instrumentation includes the Navy Precision Optical Interferometer (NPOI) and the CHARA Array, which consist of 12cm--2m aperture telescopes with sensitivity limits on the order of 6--10 mag \citep{tenBrummelaar2005,vanBelle2019}.
The Keck Interferometer \citep[e.g.,][]{swain2003,kishimoto2011} and VLTI/AMBER \citep{weigelt2012} observations with 8--10m telescopes have pushed these limits to much fainter sources ($K \simeq 10$). Increasing the sensitivity to still lower fluxes and higher spectral resolution is desirable, since only interferometry provides the precision necessary to resolve individual components in the OIR for many astrophysical systems. 

The newest frontier in OIR interferometry is fringe tracking and precision astrometry using a sufficiently bright star within a few arc-seconds of the desired target.  Without corrections, atmospheric effects cause too much
jitter in the fringes to integrate for periods long enough to detect
fainter sources.  By phase referencing, the fringes of the target can
be actively stabilized with respect to the reference source, allowing
for integration times long enough that the target can be much fainter
than the reference object.  An instrument with these capabilities is now in operation
on the Very Large Telescope Interferometer (VLTI), GRAVITY \citep{GRAVITY2017}.

GRAVITY is a second generation interferometric instrument, commissioned on the VLTI in 2016. It allows observing two objects (one bright fringe-tracking, phase reference object and one fainter science object). When GRAVITY is used with the Auxiliary Telescopes (ATs, 1.8m diameter) and the NAOMI adaptive optics system, the brighter phase reference object must have an infrared magnitude of $K \leq 8$, and the faint object has to be within 4 arcsec from the bright object but can be as faint as $K \leq 12$--13. When GRAVITY is used with the Unit Telescopes (UTs; 8m diameter), the bright object can be $K \leq 11$, and potentially even as low as $K\sim 12$ under good conditions.  The faint object has to be within 2$^{\prime\prime}$, with a current limiting magnitude around $K \leq 19$ \citep{GRAVITY2017, GRAVITY2018S2}.   The astrometric accuracy is as good as 20 $\mu$as in the best cases \citep{GRAVITY2017,GRAVITY2019S2}, with spectral differential astrometry demonstrated with a precision of 2 $\mu$as for high-mass XRBs \citep{Waisberg2017} and AGN \citep{GRAVITY2018quasar}. While the small separation between fringe-tracking and phase reference objects is a limitation for finding viable Galactic targets, we here discuss the potential new science cases particularly when considering planned upgrades allowing a separation of $\sim 30-40$".

The VLT point-spread-function for imaging is $\sim$ mas, thus the
superb stability of GRAVITY allows the determination of relative motions
in objects to precisions $\sim100$ times better than the resolution of
their structure.  GRAVITY was designed primarily to study orbital motions of stars or flare emission in the strong gravitational field of the supermassive black hole Sgr A$^*$ \citep{GRAVITY2020precession}, and has provided the most accurate distance to the Galactic centre \citep{GRAVITY2018S2,GRAVITY2018sgra,GRAVITY2019S2}, as well as the first detection of an exoplanet by OIR interferometry \citep{GRAVITY2019exoplanet}.  However GRAVITY also has the potential to revolutionise X-ray binary (XRB) studies, and in fact has already been used to study the size, structure and spectra of two Galactic high-mass XRBs, GX~301--2 and SS~433 \citep{Waisberg2017,GRAVITY2017SS433,GRAVITY2019ss433}.  While low mass XRBs are generally too small (for their distance) to be spatially resolved on mas scales using the IR band, an IR astrometric accuracy of $\sim10\mu$as could enable the first direct detection of individual contributions to the IR spectrum, as well as an independent method for obtaining system orbital parameters.  

In this paper we propose a feasibility study for how IR interferometry, using GRAVITY in particular, can be exploited to separate emission components in XRBs and thus constrain
accretion/outflow physics.  In Section~\ref{sec:science} we describe the scientific questions that IR interferometry can
help address for XRBs, particularly transient microquasars.  In Section~\ref{sec:cases} we explore the
feasibility using several typical sources as a guide, particularly for observing faint targets off-axis.   In Section~\ref{sec:maxij1820} we present the first proof-of-concept, mas scale IR interferometric observation of a Galactic transient X-ray binary, MAXI J1820+070, using GRAVITY. Finally, in
Section~\ref{sec:discuss} we summarize and make some predictions for
the coming decade of all-sky transient detections.

\section{Science motivation}\label{sec:science}

\subsection{Spatially resolving jets in XRBs}\label{sec:jets}

VLBI studies of jets in nearby AGN such as  M87 have revealed a spine and sheath geometry, as predicted by theoretical models \citep{Perlman2011,Walker2018}, as well as the collimation profile of the jet \citep[e.g.][]{AsadaNakamura2012}. Such images can be used to constrain the equations of force balance (i.e., internal versus external pressure), while variability provides information about turbulence within the flow.  M87 is in fact so close and large that the Event Horizon Telescope (global 1 mm VLBI with phased-ALMA at its core) was able to directly image the shadow of the black hole \citep[e.g.][]{EHT2019a}.

While Galactic black hole XRBs are much closer than AGN, they are typically millions to billions of times smaller, so direct imaging is a challenge and resolving the black hole shadow or inner accretion disk is well beyond the capabilities of current facilities.  With the advent of GRAVITY on the VLTI, however, there is an intriguing possibility to directly resolve expanding mas scale jets in a transient XRB outburst, and to spectrally decompose the accretion components from each other as well as from the companion star.  The advantage of XRBs compared to AGN is that one can observe millions more dynamical timescales, thus obtaining constraints on the inner disk physics over much longer relative timescales.  Direct imaging provides the best insights into jet morphology, energetics and interactions.  Because one of the pressing questions at the moment in modeling accretion flows centres on how and where particles are energised \citep[see, e.g.,][]{Romeroetal2017,BallSironiOzel2018}, the promise of pinpointing the moment when XRB jets launch and then begin to accelerate high energy particles makes them extremely valuable testbeds for constraining theory.  

\subsubsection{Compact jets and discrete ejecta}
XRBs exist as both persistent (mostly high-mass companions; HMXB) and transient (with low-mass companions; LMXB) sources, the latter of which experience periodic outburst cycles.  Within a single outburst we witness the launching and quenching of jets, in some sources repeatedly on a few-year time cycle \citep[e.g.][]{Corbeletal2013}.  Until now direct imaging has focused on radio-VLBI techniques because of the phenomenal spatial resolution, but only three compact, steady jets associated with the non-thermal-dominated `hard state' have been resolved with radio-VLBI to date: GRS~1915+105 \citep{DhawanMirabelRodriguez2000}, Cyg X-1 \citep{Stirlingetal2001} and MAXI~J1836-194 \citep{Russelletal2015}. However during state transitions to the thermal/disk-dominated `soft state' the jets transform dramatically, and increasingly more radio-VLBI studies have been able to resolve, and track the evolution of, discrete ejecta on mas scales  \citep[e.g.][]{MirabelRodriguez1994,HjellmingRupen1995,Tingayetal1995,Fenderetal1999,Mioduszewskietal2001,Miller-Jones2011,MillerJonesetal2019Nature,Brocksoppetal2013,Rushtonetal2017}.

The compact jets seen in the hard state also emit IR synchrotron emission, but it originates from too close to the black hole to be directly resolved \citep[e.g.][]{Russelletal2006,Gandhietal2011,Buxtonetal2012}.  During state transitions, discrete jet ejecta emit optically thin synchrotron from radio to IR, such that the IR flux is generally expected to be fainter than the radio (though see GRS~1915+105; \citealt{Fenderetal1997,Eikenberryetal1998}).   Despite this faintness, with the advent of  fringe-tracking and phase referencing capabilities with VLTI via the GRAVITY experiment, it may be possible to image XRB jet activity in the IR similar to what has been done with radio-VLBI.  

\subsubsection{Previous claims of OIR extended jets}
There have in fact been claims of a marginal IR detection of a compact jet of 0.2 arcsec in the source GRS~1915+105 \citep{SamsEckartSunyaev1996}, however this has never been confirmed by later detections. But recently, IR emission lines from plasma in the jets of the exotic Galactic XRB SS 433 have been spatially resolved with GRAVITY \citep{GRAVITY2017SS433}. On larger scales, extended jets of XTE J1550--564 resolved on arcsec--armin scales, detected at radio and X-ray frequencies, were almost -- but not quite -- detected by the VLT at optical wavelengths \citep{Corbeletal2002}.

\subsubsection{Detectability with GRAVITY: spatial scales}
Because GRAVITY is capable of imaging structures on spatial scales of $\sim 1$--50 mas (corresponding to, e.g., 
$\sim$0.3--20 AU for a source at 3 kpc; up to $\sim$60 AU at 8 kpc), transient features such as ejecta or jet-ISM interaction regions may be now be detectable.  Specifically, after transition to the soft state the core is no longer active but the ballistic ejecta are still moving. Based on the radio-VLBI observations, we expect these bright ejecta to be spatially resolved with GRAVITY, with motions of $\sim 10$s-$100$ mas/day (note that extremely fast motions of 100 mas/day could result in motion/smearing within a GRAVITY observation itself, depending on integration times).

The mas scale jets seen with radio VLBI are typically discrete ejecta that are themselves unresolved down to $< 1$ mas \citep[see Table 1 in][]{Miller-Jones2006}, so we do not expect them to be resolved out. One could therefore use a uv binary source model (XRB core and jet ejection, see section \ref{sec:maxij1820results} and Fig. \ref{closure}) to identify spatially resolved ejecta on scales of 1--50 mas. In the case of a nearby source with high velocity ejecta, the mas scale ejections could move on timescales as short as the exposure time. For these, if the movement is comparable to or longer than the integration time (tens of minutes), these will be detectable. A good example of this is the detection of the resolved S2 star from Sgr A$^*$, in which the dynamical timescale was shorter than the integration time, yet the uv model fitting was successful \citep{GRAVITY2018sgra,GRAVITY2019S2,GRAVITY2020precession}.

\subsubsection{Detectability with GRAVITY: fluxes}
For a typical XRB, the radio flux densities of discrete ejecta on the 1--50 mas size scales we can probe with GRAVITY are on the order of 10--500 mJy \citep[at 15 GHz, e.g.][]{FenderHomanBelloni2009,MillerJonesetal2019Nature}. Assuming a standard optically thin spectrum ($\alpha = -0.6$ to $-0.8$) would predict a $K \sim 10$--12 magnitude core and a $K \sim 15$--16 magnitude discrete ejection, meaning GRAVITY could significantly detect and resolve both components. Because such discrete ejecta would likely be several magnitudes fainter than the compact jet, and the unresolved accretion disc and (in some cases) the star,  mas-scale discrete ejecta at these fluxes would never have been noticed before in any IR data. 

\subsubsection{Detectability with GRAVITY: brief, brighter jets}
There may also be bright IR discrete ejecta that are tens of mJy
(K$\sim$10 mag), but only briefly near the start of the hard-to-soft
state transition, before fading to similar or fainter flux later in
the radio. For example, the non-spatially resolved jet flares in GRS
1915+105 were a similar flux density (in Jy) in IR and radio, with
radio occuring minutes later than IR
\citep{Fenderetal1997,Mirabeletal1998}. Both IR and radio flares
lasted $\sim 20$ minutes, strongly suggesting adiabatic losses
dominating. However during V404 Cyg's latest outburst,
\cite{Tetarenkoetal2017} found that 7 Jy sub-mm flares lasting
  tens of minutes to an hour at 666 GHz were followed by only $\sim1$
  Jy flares in the radio bands. Such flares may have very brief,
  transient IR counterparts of a few hundred mJy (K $\lesssim$ 9 mag)
  or more, lasting timescales of minutes to tens of minutes. To date
  no such IR flare has been resolved in a typical black hole transient
  (GRS~1915+105 is considered an outlier), but they may have been
  missed due to low sampling or short integration times. Such ejecta
  are likely to emit between K$\sim9-16$ during the days/week just
  after launching.  In one system, an IR flare peaking at $K < 13$ magnitude and lasting four days has been detected during state transitions, and may have been brighter on $< 1$ day timescales \citep{Buxton2004,Russell2020}. Such flares, on hour--day timescales -- if present -- will be detectable and spatially resolved if they have typical motions of $\sim 10$s-$100$ mas/day. However, we note that for very rapid flares which change flux on timescales comparable to a single observation, this could introduce image artifacts. As such, the flux variability would likely preclude all but the most basic binary model fitting in these cases.

\subsubsection{Consequences of a detection}
A clear detection would allow constraining the location, velocity, 
size/morphology, and evolution of the IR jets. Together with radio-VLBI observations, an IR detection will also provide information about the radiating particle energy distribution of the mas scale discrete ejecta. Any early-time, bright IR detections would be crucial for constraining the launch time of the jets, for comparison to X-ray timing signatures associated with this launching \citep[see, e.g.,][]{Miller-Jonesetal2012}, as they are less affected by optical depth effects. Finally, when the steady jet re-establishes itself before the outburst ends, GRAVITY could be used to investigate changes in the inner jet and movement of any interaction hotspots.

\subsection{Spectral decomposition using interferometry}\label{sec:specs}

While challenging, IR interferometry offers an exciting new dimension to the multi-wavelength studies currrently used to deconstruct the physics driving accretion and outflows.

Our understanding of XRB accretion physics has evolved significantly over the last decades, mainly due to the monitoring of entire outburst cycles in the X-ray bands via triggered instruments such as the \textit{Rossi X-ray Timing Explorer (RXTE)}, \textit{SWIFT} and more recently, \textit{NICER} \citep[e.g.][]{Bellonietal2005,MunozDariasetal2011,Mottaetal2017,Stevensetal2018}.  These observations have led to a greater understanding of the disk/jet coupling driving the discrete accretion states \citep[e.g.,][]{McClintockRemillard2006,Belloni2010} that are most pronounced in black hole XRBs, the main focus of this work. The frontier has now shifted to the lower frequency bands, as simultaneous triggering of radio and OIR observations with the X-rays has revealed a parallel evolution in the interplay between the accretion inflow in the accretion disc, jet outflows and sometimes the stellar companion.

The OIR bands in particular offer a valuable new testbed for many aspects of accretion physics in XRBs, as they can be comprised of multiple contributions from the accretion disk, jets and star, the former of which can exchange dominance during state changes.  For instance, in the soft state, the outer regions of the accretion flow can re-radiate emission from the inner zones in the OIR bands, as can the companion star itself \citep[e.g.][]{OBrienetal2002,Hynes2005,Migliarietal2007}.  In the hard state it is now well established that synchrotron emission from the jets extends into the OIR bands \citep[e.g.][]{CorbelFender2002,Russelletal2006,Russelletal2010,Buxtonetal2012,Saikia2019}, either as an extension of the flat/inverted, self-absorbed spectrum, or beyond the synchrotron self-absorption break as an optically thin power-law.   The break itself has been explicitly resolved in some observations \citep[e.g.][]{Migliarietal2006,Gandhietal2011,Russelletal2013} and recent simultaneous broadband campaigns of XRBs in outburst demonstrate that the break dynamically moves up and down in frequency through the band (\citealt{Russelletal2014}, Russell et al. subm.).

Understanding the contribution of the jets in the IR and in particular, whether the IR is above or below the break, or how the break evolves, has become a key focus of XRB studies as this places strong constraints on the jet geometry, dynamics and energetics.   For instance, recent multi-wavelength variability studies have revealed a characteristic size scale for this break \citep{Gandhi2017NatAs,Paiceetal2019}, as well as very strong near-IR to mid-IR variability from the compact jets \citep{Gandhietal2011,Baglio2018,Vincentelli2018,Malzac2018}.  Similarly the first ever IR quasi-periodic oscillations (QPOs) with a harmonic of the X-ray QPO frequency \citep{Kalamkar2016} reveal the tight coupling of jet to disk.   If the spectral slope beyond the break can be constrained, this also helps determine the particle acceleration properties and together with limits from the X-rays and $\gamma$-rays, the maximum radiative power of the jets \citep[see, e.g.][]{Laurentetal2011,Corbeletal2012,Zdziarskietal2014,Rodriguezetal2015b,Zaninetal2016, Espinasseetal2020}.  However the clear evidence for multiple contributions to the OIR  \citep[see, e.g.][]{Homanetal2005,Buxtonetal2012,Baglio2018} can make identifying the OIR jet contribution challenging.

Isolating the jet contribution to the OIR spectrum is therefore a key new milestone for understanding XRB jet physics in general, and gauging their total power budgets \citep{Corbeletal2002,Galloetal2005,HAWCSS433_2018}.  Furthermore, because of the increasing body of evidence that the accretion physics in XRBs and AGN scales predictably with mass \citep{MerloniHeinzDiMatteo2003,FalckeKoerdingMarkoff2004,McHardyetal2006,Plotkinetal2012,Koljonenetal2015,Connorsetal2017},
constraints found from IR studies of XRBs will cast light on larger scale
issues in galactic evolution such as the physics governing jet launching and power, and eventually energy released into the environment.

\section{Feasibility study for a new application of IR interferometry to
  XRBs}\label{sec:cases}

As typical XRBs have binary separations falling in the range of
1--100 $\mu$as, they are not directly resolvable with the VLTI, but
the system motion should be detectable.  The simplest, but potentially one of the most important, applications of the VLTI for XRBs is the search for a wobble in the light centroid position, which can be used to
constrain the mass function of the system.  Many systems now have
accurately measured orbital periods but the mass of the compact object
and/or companion star are, on the whole, poorly constrained. Since the
orbital separation, $a \propto P^{2/3} (M_1 + M_2)^{1/3}$ where $P$ is
the orbital period, we can therefore solve for the total system mass
with precise determinations of $a$ and $P$.

For objects with several good reference stars in the field, the VLTI can in principle be used to further constrain the
system orientation (in projection on the sky) and inclination.  The
amount of orbital wobble measured, for non-zero inclination systems,
will depend on the angle between the orbital plane of the XRB and the
reference star.  
By measuring the wobble using a number of reference stars at different angles from the
XRB, a solution for the orientation of the disc and apparent
eccentricity on the plane of the sky can be derived. While for eccentric orbits, speeds will vary from periastron to apastron, the orbits of
most XRBs (at least Roche-lobe overflow low-mass X-ray binaries) are circular, and in all cases the inclination angle of the system can also be constrained.  If the
stellar companion type and mass is known, as is the case for sources with a measured mass function in quiescence, the physical parameters of the
system can be completely characterised using this approach.

If both the orientation of the orbital plane as well as the
inclination can be constrained, a further interesting test is to
compare these with the observed orientation of the radio jets (which
are generally well known for most of the prime candidate sources considered in
this paper).  Such a test could help identify more systems with
drastically misaligned jets similar to the `microblazar' V4641 Sgr
\citep{Hjellmingetal2000,Oroszetal2001,Maccarone2002}.  The existence
of significant numbers of misaligned jets would help constrain the extent to which jet axes align more with the black hole than the outer disk, as predicted by \cite{Rees1978} and now reproduced numerically \citep[e.g.][]{Liskaetal2018}, and enable a study of timescales for which the Bardeen-Petterson effect
\citep{BardeenPetterson1975} could bring the jets back into alignment.  

\subsection{Centroid shift and potential target list} 

\begin{figure}
\centerline{\includegraphics[width=0.48\textwidth]{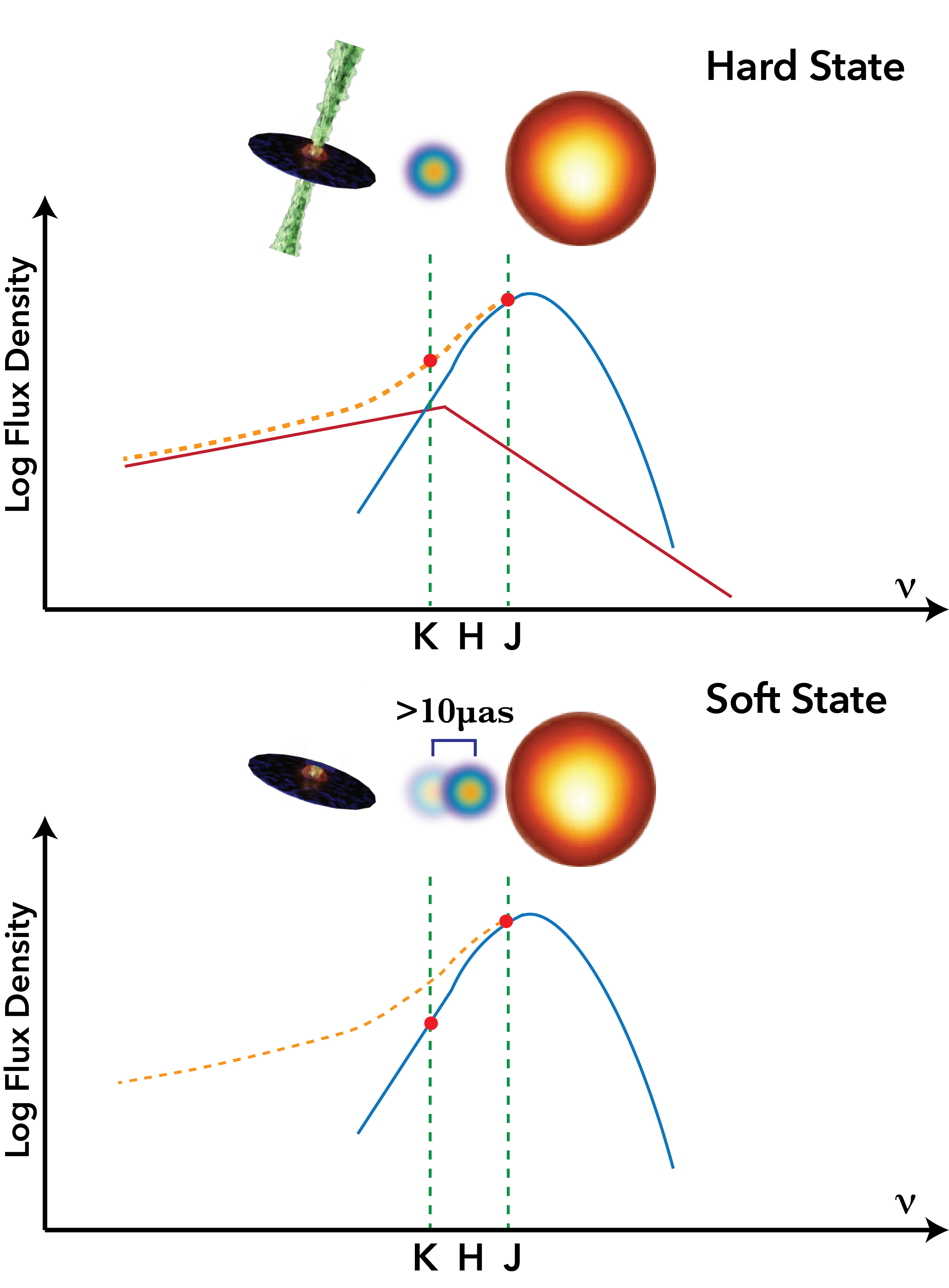}}
\caption{Schematic of the centroid shift expected during state
    changes, due to the disappearance of the jet IR synchrotron
    emission. \textit{Top:} System in the hard state of a LMXB, where both jet
    (red broken power law) and companion star (blue blackbody) contribute comparably to the flux density in the IR bands (orange dashed).
    \textit{Bottom: }  Same system, at the same orbital phase, after
    transitioning to the soft state and jet contribution to the IR is
    quenched, resulting in a shift of the image centroid towards the
    companion star in the K-band. The orbital wobble, most prominent when the star is producing the IR emission, is largely reduced when the jet is dominating.  Note that another centroid shift could be potentially seen between K and J bands during state transitions.}
\label{centroid}
\end{figure}

The most challenging but potentially most exciting detection beyond orbital effect would be a shift in the image centroid as a function of accretion state,
driven by waxing and waning in the three main system components in the
IR band.  Understanding the orbital period via the wobble technique or
otherwise will be a necessary first step, to allow comparison of the
source at the same orbital phase    in different states, and
potentially to co-add images from several orbits.  A schematic of how
this technique would work to identify the jet contribution can be seen
in Figure~\ref{centroid}.  If the system is observed at the same
orbital phase during the hard and soft states respectively, and the binary separation is
sufficiently large, the shift in centroid position should be
detectable for a few systems with GRAVITY.  If
significant IR emission is thought to be contributed from
irradiation in the star, this technique together with spectral
modeling would also help to break the degeneracy between that and disk/jet, if comparisions are made between state changes.

For both high- and low-mass XRBs,
astrometry with the current specifications of GRAVITY on the VLTI should be achievable if the target has $K\lesssim 16$ and another
$K\lesssim 11$ magnitude star lies within 2 arcsec from
the target, or vice versa (brighter target, fainter reference star).  For our specific interests, transient LMXBs (or bright flares in persistent LMXBs) are preferable, but in Table 1 we provide a list of all the best known sources which would be good candidates based  on the best known estimates of distance, orbital period and component masses from the literature to calculate the orbital separation.  Starting from the 40 XRBs with radio detections (i.e. evidence of jets), we find 15 that have apparent orbital separations on the sky $a > 10 \mu$as (in descending order of $a$).  We include sources too far north for the VLT such as the high-mass X-ray binary (HMXB) Cyg X--1, because it is a canonical object with well-constrained physical parameters that could be useful for future interferometry instruments \citep[on a northern interferometer such as the Large Binocular Telescope, the CHARA Michelson Array and the Magdalena Ridge Observatory Interferometer (MROI);][]{Angeletal1998,tenBrummelaar2005,Buscher2013,Gies2019}.

These are mostly LMXBs (including black hole and neutron star sources)
and also some HMXBs with radio emission. We consider these the current
best targets for VLTI attempts with GRAVITY (except for three sources
that are too far north, shown in italics in the table); we do not
include the other systems in the table as they have smaller apparent
orbital separations on the plane of the sky, and are more challenging
for GRAVITY. The observed (not de-reddened) $K$-band magnitudes are
also tabulated, as are the number of $K < 12$ stars within
40$^{\prime\prime}$ of the X-ray binary, listed in the 2MASS
catalogue.  The reason this list extends beyond GRAVITY's current
beam-forming capabilities is because of the proposed enhanced
sensitivity and the enlarged field of view of a possible GRAVITY+
upgrade (in prep.), that was presented at the ``The Very Large
Telescope in 2030'' conference, ESO Garching, June 17-20, 2019.  All
of these objects also have radio flux densities (not shown) which can
be used to extrapolate a first order estimate of the expected IR flux
from the jets. In some cases the jet contribution to the $K$-band IR
emission has been estimated in the hard state; for these we estimate
the phase shift of the centroid between jet on and jet off (if instead
the star dominates) states.  If instead the shift is from jet to accretion disc over the transition, there would be no phase shift in the XRB during such a transition (this is most likely the case for the LMXBs with the faintest companions, such as GX 339--4, XTE J1550--564 and Cen X--4) and so the phase shift calculated from the ratio $\frac{I_j}{I_j + I_s}$ (see below) represents an upper limit in these systems. However in many of the sources in Table 1, the star is large and brighter than (or of comparable brightness to) the accretion flow \citep[GX 301--2, CI CAM, Cyg X--1, SS 433, GRS 1915+105, V4641 Sgr and GRO J1655--40 as shown in Fig. \ref{centroid}; e.g.][]{Kaper1995,Miroshnichenko2002,Migliarietal2007,Hillwig2008,vanOersetal2010,Rahoui2011,MacDonaldetal2014,RussellShahbaz2014} and so a phase shift is expected.

There are currently at least six known systems with orbital separations $a > 50 \mu$as, which would yield a $5\sigma$ detection with GRAVITY of an astrometric
shift over the orbital period. This shift also depends on the mass ratio; for systems in which the black hole mass is much greater than the companion mass, the orbital wobble of the companion could be as large as twice the orbital separation. Most orbital periods of the sources in
Table 1 are days to weeks. Exoplanets, by comparison have periods on the order of years, so
instrument drifts and systematics over the longer time frame introduce
additional astrometric errors which are not relevant for
microquasars.

Unfortunately, none of the targets in Table 1 have known nearby stars within 2$^{\prime\prime}$ in 2MASS, with the closest being GRS 1915+105 with a $K = 13.2$ star $3.3^{\prime\prime}$ from the target. However, a faint star close to the brighter XRB would not be easily detectable in the 2MASS survey, so it is possible some close stars have been missed by 2MASS. We did also check higher resolution images from VISTA surveys \citep[][including VVV and VHS]{Minnitietal2010} for nearby stars.  If off-axis capabilities allow a phase reference star within $\sim 30$--40 arcsec (as suggested in a GRAVITY+ white paper; Gravity Collaboration et al., in preparation), this dramatically increases the feasibility. Most targets have several (up to 17) bright $K < 12$ 2MASS stars within $40^{\prime\prime}$ (Table 1). GX 301--2 and CI Cam have the widest angular orbital
separations, but in both these systems the jets never brighten to an
IR flux level comparable to that of the companion star. Some of the
other sources do have relatively bright jets though, and we
investigate these further below.


\begin{table*}
\begin{center}
\caption{The XRBs (with radio detections) possessing the widest orbital separations on the sky$^1$ (all with $a > 10 \mu as$). The best interferometry candidates are those that are expected to have a shift of the phase of the centroid due to a changing IR jet contribution (final two columns).}
\begin{tabular}{llllllll}
\hline
Source$^2$ & BH/NS$^3$ & K mag& $a$ ($\mu as$)$^4$ & $K<12$ stars$^5$ within 40$^{\prime\prime}$ & Refs & $\frac{I_j}{I_j + I_s}$ & $\delta\phi$ \\
\hline
\textbf{GX 301--2 (BP Cru)} & NS & 5.7   	 & $243 \pm 15$  & 3 stars, $K = 6.7$--10.5 & 1--2 & -- & -- \\
\emph{CI Cam} & ? & 4.1--4.7  & $223 \pm 176$ & 1 star, $K = 10.2$ & 3 & -- & -- \\
\emph{Cyg X--1} & BH & 6.5	 & $127 \pm 16$  & 1 star, $K = 8.9$ & 4 & 0.006 & $(0.07 \pm 0.01)^\circ$ \\
\textbf{SS 433} & BH? & 8.2	 & $75 \pm 7$	 & 0 stars & 5--6 & -- & -- \\
\emph{V404 Cyg} & BH & 7.7--12.5 & $61 \pm 5$	 & 4 stars, $K = 9.5$--11.8 & 7 & $>0.80$  &$(4.8 \pm 0.9)^\circ$ \\
GRS 1915+105 & BH & 11.4--13.5& $55 \pm 14$   & 6 stars, $K = 9.0$--12.0 & 8 & 0.05--0.2 &$(0.68 \pm 0.51)^\circ$ \\
~~~Flaring state & & & & & & $\sim 1$ & $(4.7 \pm 1.2)^\circ$ \\
GX 13+1 & NS & 11.9--12.6 & $44 \pm 8$ & 17 stars, $K = 6.9$--12.0 & 9--10 & -- & -- \\
Cir X--1 & NS & 7.2--11.9 & $28 \pm 11$ & 10 stars, $K = 9.8$--11.9 & 11--14 & -- & -- \\
GRO J1655--40 & BH & 11.0--13.3& $25 \pm 3$ & 10 stars, $K = 9.1$--12.0 & 15--16 & 0.25 & $(0.53 \pm 0.05)^\circ$ \\
A0620--00 & BH & 9.9--14.5 & $17 \pm 2$	 & 0 stars & 17--18 & -- & -- \\
Cen X--4 & NS & $<13.0$--14.8 & $16 \pm 5$	 & 1 star, $K = 10.8$ & 19--21 & -- & -- \\
V4641 Sgr & BH & 12.7--13.7 & $13 \pm 2$	 & 4 stars, $K = 8.3$--10.9 & 22--23 & $\gtrsim 0.9$ & $(1.1 \pm 0.2)^\circ$ \\
XTE J1550--564 & BH &13.0--17.4& $13 \pm 2$	 & 7 stars, $K = 9.3$--12.0 & 24 & 0.9 &$(1.0 \pm 0.2)^\circ$ \\
GRO J1719-24 & BH & $<13.5$--18.3 & $>12 \pm 5$	 & 4 stars, $K = 8.0$--11.9 & 25--26 & -- & -- \\
\textbf{MAXI J1820+070} & BH &9.5--15.1& $>9 \pm 1$ & 1 star, $K = 12.0$ & 27,28 & $>0.5$ & $> 0.4^\circ$ \\
\hline
\end{tabular}
\normalsize
\end{center}
$^1$We do not include $\gamma$-ray binaries in this table, which are not thought to have extended jets.
$^2$Sources in \textbf{bold} have been observed with GRAVITY on VLTI; sources in \emph{italics} are too far north for the VLTI (Declination $> +25^\circ$).
$^3$BH = black hole; NS = neutron star.
$^4$Errors on the estimated values of orbital separation are propagated from the errors in $d$, $P$, $M_1$ and $M_2$ (if the mass of the neutron star is not known we adopt $M_2 = 1.4 \pm 0.6 M_{\odot}$). For GRO J1655--40 the value of $a$ is given for distance $3.2\pm 0.3$
kpc \citep{Gandhi2019GAIA}.
$^5$The number of $K < 12$ stars within 40$^{\prime\prime}$ of the X-ray binary, and their range of magnitudes (data from the 2MASS catalogue). The $K$-band jet contributions in the hard state are estimated from (in some cases modelling of) spectral energy distributions in \cite{Fenderetal2000,Fenderetal1997,Russelletal2006,vanOersetal2010,Migliarietal2007,Russelletal2010,Russelletal2013,Russell2018ATel11533}. The references for the distances, orbital periods and masses can be found in the following articles and references therein:
(1) \cite{TomsickMuterspaugh2010};
(2) \cite{Doroshenkoetal2010};
(3) \cite{Thureauetal2009};
(4) \cite{Oroszetal2011b};
(5) \cite*{BlundellSchmidtobreickTrushkin2011};
(6) \cite{Lopezetal2006};
(7) \cite*{KharghariaFroningRobinson2010};
(8) \cite{vanOersetal2010};
(9) \cite{Corbet2003};
(10) \cite{Corbetetal2010};
(11) \cite*{Clarksonetal2004};
(12) \cite{JonkerNelemans2004};
(13) \cite{Toroketal2010};
(14) \cite*{Jonkeretal2007};
(15) \cite{Greeneetal2001};
(16) \cite{Gandhi2019GAIA};
(17) \cite{GonzalezHernandezCasares2010};
(18) \cite{Cantrelletal2010};
(19) \cite*{Shahbazetal2014};
(20) \cite{Chevalieretal1989};
(21) \cite{Hammersteinetal2018};
(22) \cite{Oroszetal2001};
(23) \cite{MacDonaldetal2014};
(24) \cite{Oroszetal2011a};
(25) \cite{Masettietal1996};
(26) \cite{dellaValleetal1994};
(27) \cite{Torresetal2019};
(28) \cite{Atri2020}.
\end{table*}

The spacing of an interferometer's fringes on the sky is $\lambda/B$ (where $B$ is the baseline length),
analogous to the $\lambda/D$ resolution of a single optical telescope.  The
phase of the centroid of the compact object (dominated by either the
jet or the disk, we will use the jet as an example below) and
companion star can be calculated from:
\begin{equation}
\phi = 360^\circ \frac{I_j}{I_j + I_s} \left(\frac{2 \pi a_{\rm r}}{\lambda/B}\right),
\end{equation}
where $a_{\rm r}$ is the binary separation in radians and $I_j$ and
$I_s$ are the measured jet and star flux at a given frequency,
respectively.  This equation assumes (a) $2\pi a_{\rm r} \ll \lambda/B$ (the marginally resolved limit), (b) the angle between the line connecting
the XRB and guide star on the sky and the projection on the sky of the
line connecting the two telescopes, is zero (this geometry gives the
highest resolution; $\cos(\rm{angle}) = 1$), and (c) $\sin ~ a_{\rm r}
\approx a_{\rm r}$, which is valid for the small angles we are dealing
with here. The ratio $\lambda/B$ for $K$-band is 4.37 mas (we adopt a baseline of 110 m for the VLTI).  For an XRB
with an orbital separation on the sky of $a_{\rm r} = 100$ $\mu$as,
the maximum phase shift between the hard and soft states (due to jet
quenching) is on the order of $\delta\phi\sim8^\circ$, which
corresponds to $\sim 100~\mu$as on the sky (i.e., the orbital
separation), and thus should be detectable with GRAVITY. This maximum shift scenario, in which the jet or disc produces $\sim 100$ per cent of the flux in the hard state and the star produces $\sim 100$ per cent of the flux in the soft state, will generally not be the case in reality. We therefore estimate the phase shift using the ratio $\frac{I_j}{I_j + I_s}$ and these are given in Table 1. The estimates of this ratio are taken from the literature using the relative jet and star contributions at the frequency of $K$-band \citep{Fenderetal1997,Migliarietal2007,vanOersetal2010,Russelletal2010,Chaty2011,Russelletal2013,Russell2018ATel11533,RussellShahbaz2014,Bernardini2016,Maitraetal2017}.

\begin{figure}
\centerline{\includegraphics[width=0.69\textwidth]{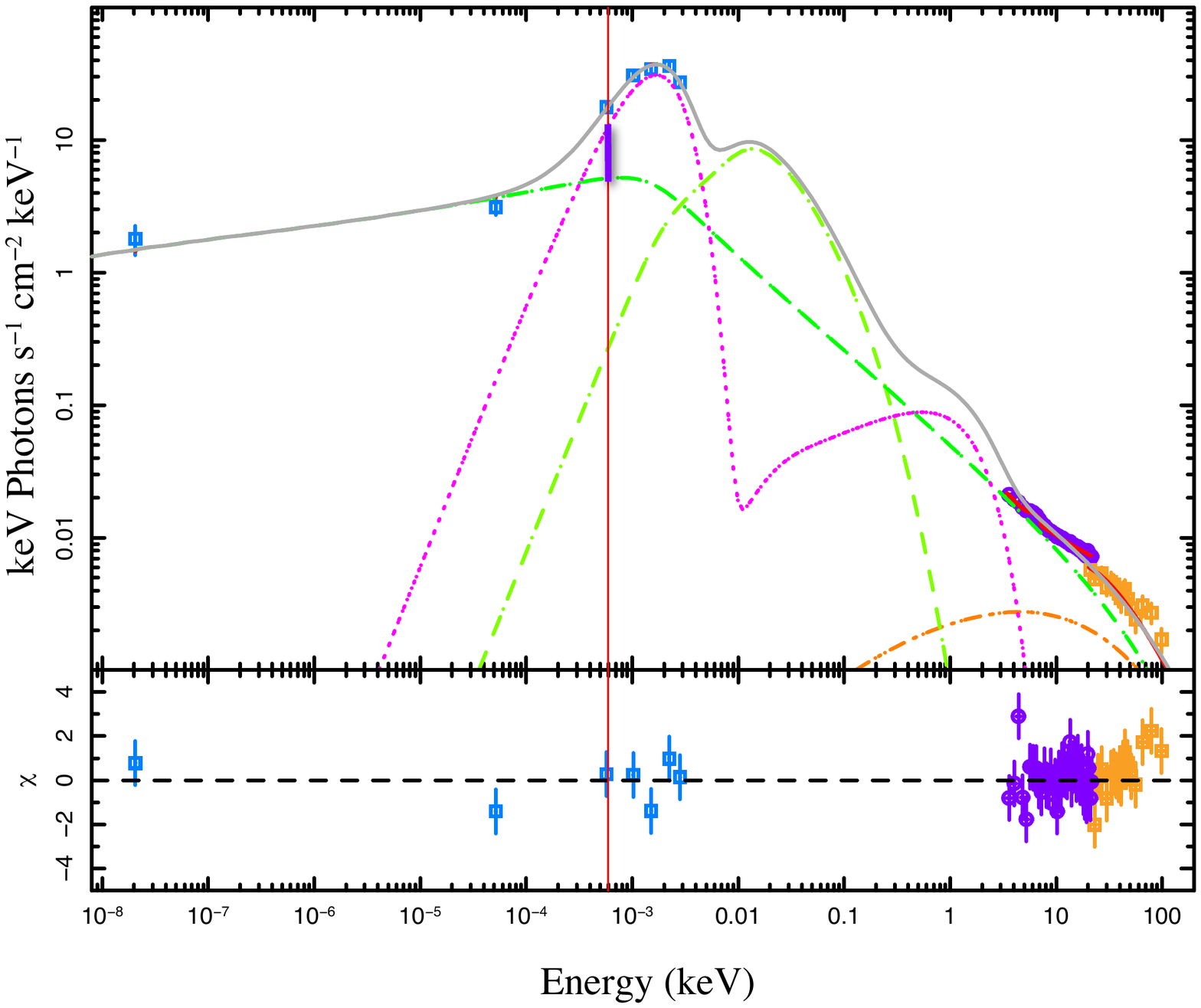}}
\caption{An example simultaneous, broadband spectrum from GRO~J1655-40
  in the hard state, from \citet{Migliarietal2007}, illustrating the K
  band (vertical red line) and the difference in flux between the jet
  synchrotron emission (dashed green line) and thermal companion star (dashed purple line; note the accretion disk is also present but very distinct at higher energy). }
\label{j1655}
\end{figure}

To test the feasibility of this new class of measurement, we use some
examples of known Galactic XRBs where the binary separation and
broadband spectral energy distribution are well constrained, allowing
an estimation of the potential centroid shift.  In Figure ~\ref{j1655} we show
an example simultaneous, broadband spectrum from the Galactic
transient GRO~J1655-40, which we have chosen because of its
large orbital separation, the fact that in the hard state
the jet flux is a reasonable fraction of the $K$-band flux, and that jet
models have been fitted to this data set \citep[][note that the conclusions would not be affected if the jet synchrotron cuts off well before the X-ray band]{Migliarietal2007}.
This source is visible to the VLT though it has currently returned to
quiescence, but during outburst had a $K$-band magnitude of 11.
We have
calculated the predicted shift 
assuming the jet is
entirely quenched in the $K$-band during the soft state \citep[this is
supported by the dramatic quenching of radio and mid-IR flux in the
soft state of GRO~J1655-40;][]{Migliarietal2007}.  Based on the
spectrum, the jet contributes $\sim25\%$ of the total flux in the hard
state, leading to a predicted shift in phase of
$\delta\phi=0.5^\circ$, which is in the
detectable range by GRAVITY during a future outburst.  There are ten
field stars brighter than $K = 12$ within 40" of GRO~J1655-40, but none within $2^{\prime\prime}$. If the field of view for phase referencing could be increased to >30", a full orbital solution would be possible for this source.  

The `persistent transient' GRS~1915+105 is another interesting potential target, having been in an outburst state associated periodically with jet ejections for the last 20+ years. During these flares the jet produces almost all of the $K$-band flux, and the estimated shift in phase of the centroid is
$\delta\phi \sim 5^\circ$.

We should also consider spectro-astrometry. GRAVITY provides a spectral range between 2.00--2.45 $\mu$m. The astrometric shift between blue $K$-band (larger star contribution, in the case of GRO J1655--40) and red $K$-band (dominated more by the jet) is larger than the astrometric shift of the average $K$-band from photometry. It may be possible to distinguish features in the spectrum from regions dominated by the jet, disc and companion star. For example, a strong Br-$\gamma$ line is expected from the accretion disc, various absorption lines from the companion star, and the red part of the $K$-band continuum dominated by the jet. If so, this would also give an interesting astrometric signal of different components at different positions. This has been recently achieved for SS 433, in which emission lines from the jets were found to be spatially offset from the continuum, the jets resolved at $\sim 1$--10 mas scales \citep{GRAVITY2017SS433,GRAVITY2019ss433b}, but this is the only known source with optical and IR emission lines from the jets.

In the last two columns of Table 1 we show the fractional jet contribution, and calculate the centroid phase shift, based on spectra taken from the literature for several of the best candidates. The errors on the phase are propagated from the
errors in $a$ and (for three sources) the jet
contribution (if it is poorly constrained). Cyg X--1 has the widest orbital separation on the sky
for sources with a jet contribution (making it a prime target for measuring the orbital wobble), but since
it is a HMXB its IR flux is dominated by the companion, and the jet
contributes just $\sim 0.6$\% of the $K$-band flux at most. The predicted shift in
the phase of the centroid due to the jet quenching is therefore small
for Cyg X-1.  In contrast, the IR jet flux of V404 Cyg was at least $\sim 80$\% in the hard state
during its outburst in 1989 \citep[and possibly during the flares during the 2015 outburst; see e.g.][]{Maitraetal2017}, and the resulting phase shift is large; $\sim
5^\circ$. Both Cyg X--1 and V404 Cyg are located too far north to be
observable from the VLTI, however it is likely that XRBs with similar
orbital parameters and jet contributions (to the latter transient source at least) will be discovered in the
coming years.
A recent example is MAXI J1820+070, which was discovered in 2018 and has an orbital separation of $a > 10 \mu$as (Table 1). We discuss the first GRAVITY observation of this source in the next section.  A good target, observable from VLTI, is GRS~1915+105. During its flaring state the jet produces almost all of the IR emission, so we predict a centroid phase shift of
$\delta\phi \sim 5^\circ$. The three remaining sources with measured IR jet contributions have predicted phase shifts due to a changing jet contribution of $\sim 0.3$--$1^\circ$, making them the next best
currently known sources visible from the VLTI.

\subsection{Accounting for parallax and proper motion}\label{sec:dontkillthetechnique}

This technique requires measuring changes in the centroid at the 10 $\mu$as level, on timescales of hours--days (for abrupt state changes) to weeks--months (orbital modulation, co-adding images from several orbits). Parallax and proper motion will therefore be significant, and their effects will need to be removed, before centroid shifts from orbital motion and state changes can be detected with sufficient accuracy \citep[see e.g.,][for discussions on this related to X-ray binaries]{TomsickMuterspaugh2010,Atrietal2019}. The targets in Table 1 with estimated centroid phase shifts due to a changing IR jet contribution have known parallax and proper motions measured from Gaia and/or radio VLBI \citep[][and references therein]{Gandhi2019GAIA,Atri2020}. The uncertainties on these measurements are $\pm$ 0.02--0.11 mas for the parallax and $\pm$ 0.05--0.22 mas yr$^{-1}$ for the proper motion (all except XTE J1550--564 have measurements). By the time of the GRAVITY upgrade, these uncertainties will very likely be smaller, with updated values from subsequent Gaia data releases, and new VLBI observations.

Using these existing astrometric solutions, the positional uncertainties for any given epoch are of order 0.11--0.60 $\mu$as due to uncertainties in the parallax and 0.14--0.60 $\mu$as d$^{-1}$ due to uncertainties in the proper motion (for parallax this will vary depending on location of the target on the sky and the time of year). Star spots on the surface of the companion could also produce light centroid jitter, with a maximum centroid shift of the order of a few $\mu$-AU, with only extreme cases from superflares of some stars producing shifts up to $\sim 100$ $\mu$-AU \citep[e.g.][]{Morris2018}, or 0.02 R$_{\odot}$, or 0.1 $\mu$as at a distance of 1 kpc. This effect of star spots is much smaller than the other uncertainties discussed here. The hot spot/stream impact point can make a small contribution to the optical emission in quiescence \citep{Cherepashchuk2019}. The hot spot is only significant in some LMXBs (not HMXBs) in quiescence as an additional thermal emitter to the optical continuum, and emission lines, but in outburst these are negligible. Since we are interested in the K-band continuum, and the disc dominates the thermal emission in outburst, the hot spot will play an insignificant role in the K-band continuum. Considering these arguments, an abrupt jet contribution change on day timescales at the $> 10 \mu$as  level will be easily detectable over the smoother changes in target position due to parallax and proper motion. Over a period of a month, the uncertainties grow to 3--18 $\mu$as and 4--18 $\mu$as, which becomes significant for measuring orbital and state changes on the 10 $\mu$as level. However, with sufficiently accurate measurements from Gaia and VLBI, it will be possible to remove these effects.  In addition, if enough GRAVITY measurements are made over year timescales, it may be possible, to independently measure parallax and proper motion using GRAVITY. This would be extremely interesting for constraining the distances, Galactic distribution, natal kicks and origins of the systems \citep[e.g.][]{Mirabeletal2001,MillerJones2014,Atrietal2019}.

Centroid shifts on the 10 $\mu$as level also require these target phase shifts to be measured relative to a phase calibrator source. The reference source is typically a star, which has its own parallax and proper motion. One will therefore need to determine the relative parallax and proper motion signatures between the reference star and the target. For the comparison star 2MASS J19053212--0016155 used below with the observation of MAXI J1820+070, we see that it has parallax and proper motion measured by Gaia DR2, with uncertainties of 0.051 mas yr$^{-1}$ and 0.076 mas yr$^{-1}$, respectively. These are very similar to our targets above, and will also be improved with future Gaia releases. One may choose to select reference stars with smaller uncertainties; perhaps more distant stars. For sufficiently high precision astrometry to measure orbital shifts, the systematic uncertainties related to the calibrator throw from the reference star should also be known.  While these are believed to be negligible for the current few-arcsecond calibrator throw, they may become important for a wider-field GRAVITY upgrade. We note that for many past and current GRAVITY science cases, true off-axis astrometry to a reference source has not been required, since all extragalactic targets are typically much further than our targets and therefore have negligible proper motion and parallax.

\section{First GRAVITY detection of a Galactic transient XRB}\label{sec:maxij1820}

As an initial proof-of-concept we here present the first triggered observation of a transient X-ray binary with GRAVITY. MAXI J1820+070 is a new black hole candidate XRB that was first detected in March 2018 by the MAXI all sky X-ray monitor and ASAS-SN optical transient survey \citep{Tuckeretal2018,Kawamuro2018,Denisenko2018}.  During its outburst rise it became one of the brightest XRB transients to date, becoming the second brightest X-ray source on the sky after Sco X-1, likely owing to its proximity \citep[2.96 $\pm$ 0.33 kpc, recently measured from radio parallax;][]{Atri2020}. Unlike most outbursts, which usually transition from the steady jet-dominated state (hard X-ray state) to the ballistic jet state (transition through the intermediate states), MAXI J1820+070 rose very quickly to its maximum brightness and stayed there for several months, remaining in the steady jet state, meaning that we had an excellent chance to observe a very bright nearby system, with $K$-band magnitude of $\geq 9.5$ \citep{Mandal2018}. A flux density of 300 mJy was measured in the mid-IR from VISIR on VLT UT3 \citep{Russell2018ATel11533}.

We observed MAXI J1820+070 with GRAVITY on the VLT Interferometer, using all four UTs (DDT 2101.D-0517) on the night of 2018 May 31 -- June 1. The X-ray binary was observed at 04:28--05:45 UT and 06:11--06:59 UT on June 1. A comparison star 2MASS J19053212--0016155 (HD 177631) was observed at 05:47--06:10 and 07:01--07:31. All observations were performed under photometric conditions, with a variable seeing of 0.6--0.9$^{\prime\prime}$. We first closed the loop of the MACAO visible adaptive optics systems on target with each telescope. We observed in low spectral resolution, placing the science fiber away from the target in order to increase flux in the fringe tracking fiber. 

We acquired one (55 mins on source, with 40 mins of science exposure) GRAVITY observation of MAXI J1820+070, when a steady, compact jet was being launched. This observation was a success in terms of feasibility, working remarkably well on a technical level. Despite the source being faint, fringe tracking \citep{lacour2019} was possible for $\gtrsim 50\%$ of the total observation time, providing good data quality (for the comparison star, fringes found in all baselines). The data were reduced with the standard GRAVITY pipeline using the default settings \citep{lapeyrere2014}.

\begin{figure}
\centerline{\includegraphics[width=0.48\textwidth]{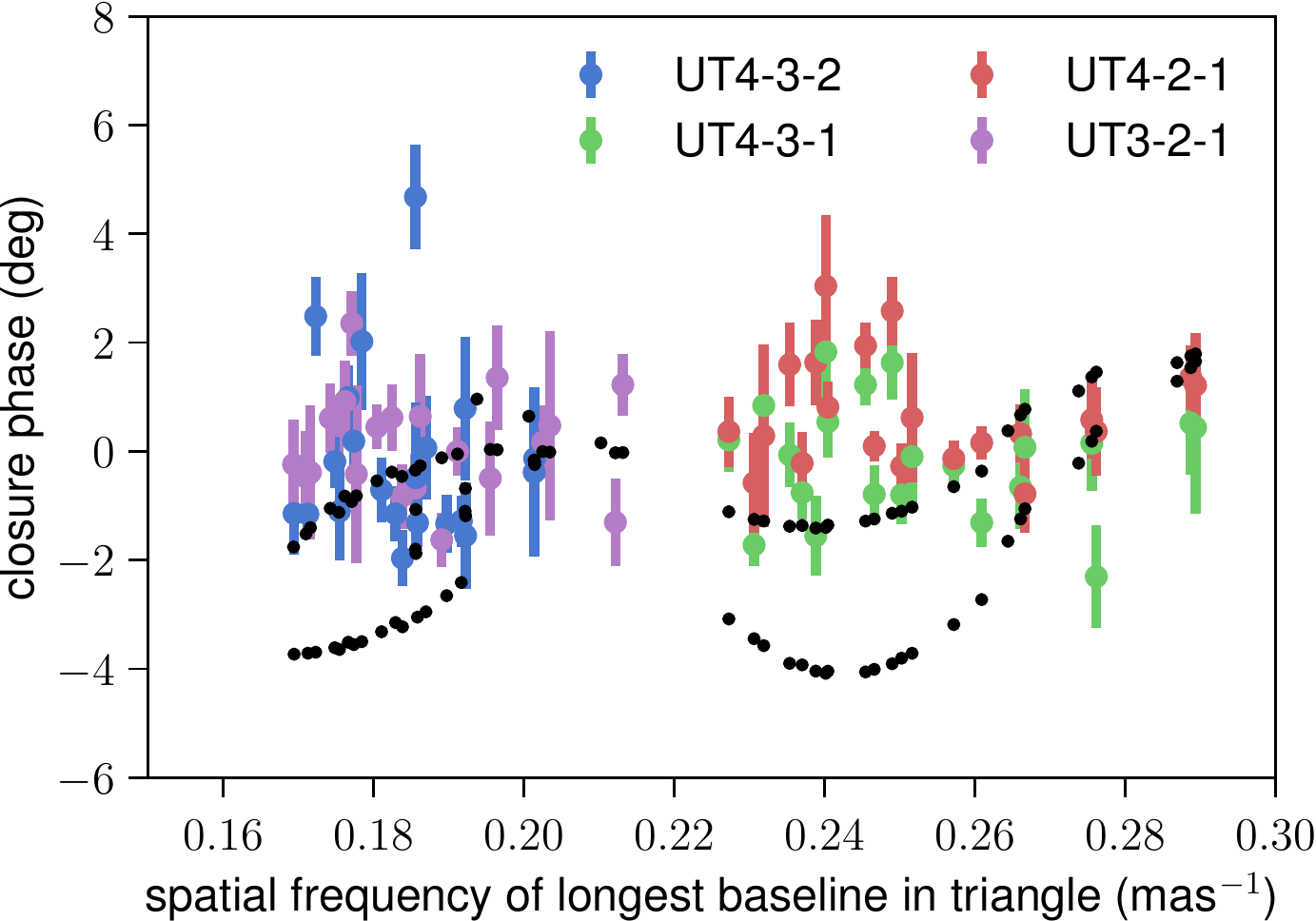}}
\centerline{\includegraphics[width=0.4\textwidth]{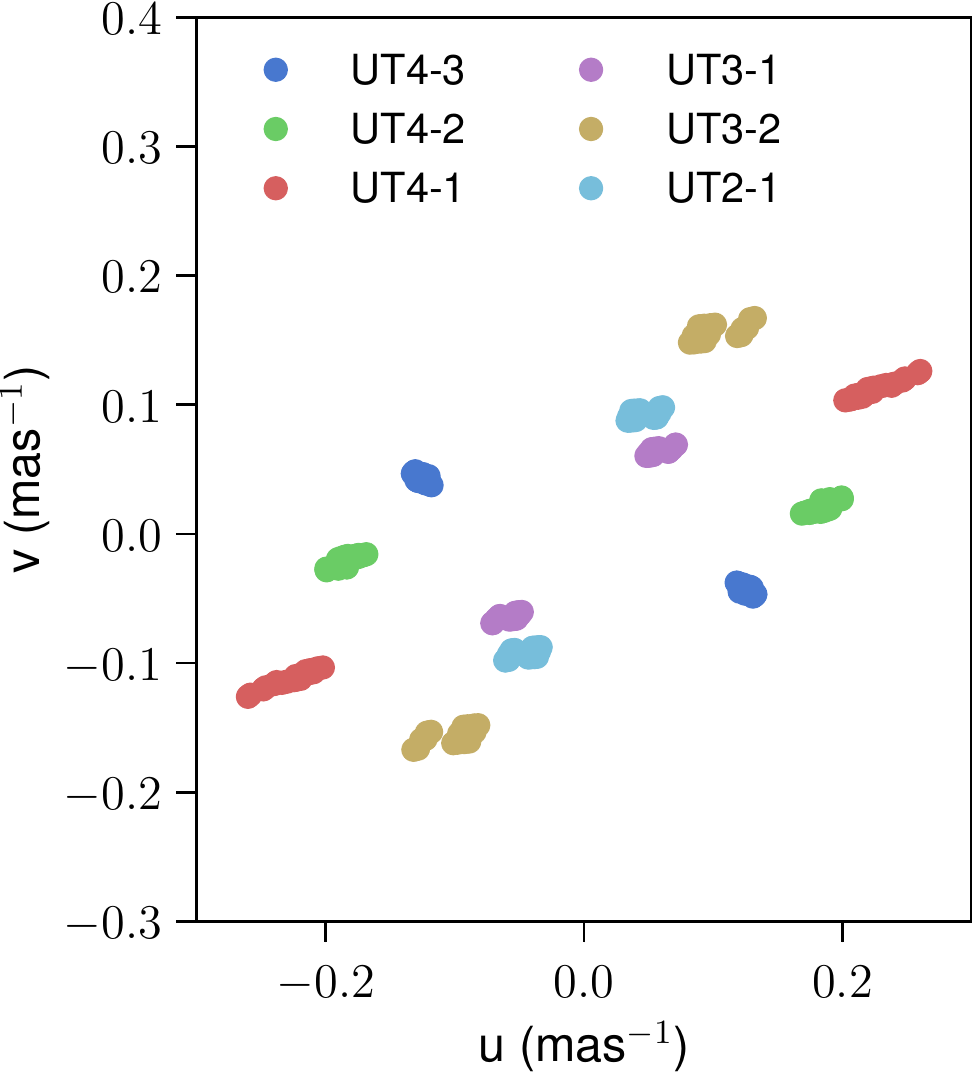}}
\caption{Measured closure phases (\textit{top}) and uv-coverage (\textit{bottom}) from our GRAVITY observation of MAXI J1820+070, colored by baseline triangle or individual baseline. In the upper panel, colored points with error bars are compared with the prediction of a binary model with a flux ratio of 0.03 and a separation of 10 mas at PA of 45 deg E of N (black dots). The measured closure phases constrain the flux ratio for any secondary ejected component to be 0.01--0.1 for separations 1--50 mas over all position angles. A single unresolved component would have closure phase $= 0$ deg at all spatial frequencies; our measured closure phases have median and rms $0.1 \pm 1.5$ deg, consistent with zero. The data are of sufficient quality to detect a faint, offset second component if present for a future transient.}
\label{closure}
\end{figure}

\subsection{Results}\label{sec:maxij1820results}

The GRAVITY data show calibrated squared visibilities ($V^2$, squared correlated flux normalized to the value at zero baseline) consistent with a constant value of $\simeq 0.8$--0.85. There is no apparent drop with increasing spatial frequency, i.e. the source is unresolved. We obtain an upper limit to the source size by fitting a Gaussian source model separately to each exposure. We allow for a variable zero-baseline visibility level to allow for coherence loss. The measured sizes are very small, with an upper limit of Gaussian FWHM $\lesssim 0.1$ mas. We measure closure phases on all triangles (Fig. \ref{closure}, upper panel) which are consistent with 0 with an rms of $\simeq 1$ deg. Closure phases of zero are expected for an unresolved source \citep{lachaume2003}.

A secondary component would show up as an oscillatory signal in the closure phases and from a limit $\lesssim$ 2 deg the flux ratio is $\lesssim$ 2\% for separations $\gtrsim$ 1 mas. This argues for an unresolved or marginally resolved source, no asymmetry, and possibly some additional (over resolved/extended) background flux, which reduces the visibility at short baselines. 

The photometric flux is  a factor $\sim 5.9$ lower than the calibrator ($K=10.0$), which implies the source magnitude was $K \sim 11.9$ at the time of observation. The most likely model for the source is an unresolved point source, and some extended background. In the acquisition camera, the $H$-band flux of the calibrator is a factor of $\sim 3.6$ larger, which would correspond to $H \sim 11.4$. 
The unresolved core is fully consistent with expectations from jet models, as the IR synchrotron emission is likely dominated by regions within $10^3-10^4$ $r_g$, where $r_g = \frac{GM}{c^2}$ is the gravitational radius of the black hole \citep{Gandhietal2011,Gandhi2017NatAs}.
Nevertheless, there are no bright jet--ISM interaction sites, constraining the dissipation of energy associated with the steady jet on these scales. This observation was made during the prolonged hard state during the first part of the outburst of MAXI J1820+070. Later in the outburst, the source made a transition to the soft state, but unfortunately GRAVITY was unavailable at the time, and no observations of discrete ejecta were possible during this outburst.

Our size constraint is the the tightest direct (i.e. from imaging) limit on the size of the NIR emitting region in a hard state XRB. The source size limit of $\lesssim 0.1$ mas corresponds to a distance of $4.4 \times 10^7$ km, or $\sim5 \times 10^6$ $r_g$ for a 5 M$_{\odot}$ black hole at 3.0 kpc \citep[][5 M$_{\odot}$ is a lower limit; this source size estimate reduces for higher BH masses]{Torresetal2019,Atri2020}.
While this size determination is less constraining than indirect methods, it agrees with those inferred from the variability timescales \citep[e.g.][]{Casellaetal2010,Kalamkar2016,Gandhi2017NatAs} and model predictions of XRB jet spectra \citep*[e.g.][]{MarkoffFalckeFender2001,MarkoffNowakWilms2005}. 
The 0.5 mas size scale of the extended jet at radio wavelengths (Miller-Jones et al. in prep) is comparable to the resolution of the GRAVITY observation at IR wavelengths, however the size scale of the steady, compact jet is expected to be orders of magnitude smaller at IR wavelengths. Nevertheless, this jet will be expanding and pushing into the surrounding ISM, and the fact that it was in the bright and steady jet state for two months means that the interaction zone with the ISM likely moved out to large scales. Conservatively estimating a velocity of the jet-head to be 0.01c, this would be $\sim 10^{15}$ cm or $\sim 10^{-3}$ pc.  At a distance of $3.0$ kpc this length-scale corresponds to 10s of mas, well within the capability of GRAVITY.  Our measurement is thus an excellent feasibility study and clearly demonstrates that spatially resolving jet ejecta and jet-ISM interaction regions on scales of 1--50 mas ($\sim$0.3--20 AU for a source at 3 kpc; up to $\sim$60 AU at 8 kpc) is possible with GRAVITY.

\section{Discussion}\label{sec:discuss}

The case studies presented here demonstrate the untapped potential of
upcoming high-precision OIR interferometry instruments such as GRAVITY
to open a new discovery space for sources other than their intended
targets.  It is extremely timely to begin tests of this technique
during commissioning of upgrades to GRAVITY for example, because current high-energy wide field monitors
(e.g. \textit{Swift}, \textit{Fermi}, \textit{MAXI}) are being joined
by deeper X-ray all-sky instruments such as eRosita \citep{Merlonietal2012}, and the first generation of mid-to-high frequency range radio all-sky monitors (e.g. MeerKAT; \citealt{Fenderetal2017}).  We expect tens of new XRBs to be
discovered in outburst, most of which will likely be in the
southern hemisphere and hence visible from the VLT.  Well constrained
orbital parameters for as many of these systems as possible will be
vital for constraining the nature of the various compact primaries, as
well as binary evolution models in general.

In the future, other more challenging applications can be considered.
For instance, wavelength-dependent centroid shifts may be detectable
when the contributions of two components in a single state are
oppositely decreasing/increasing very quickly in the IR band.  For
instance, in the hard to soft state transition, the jet is likely
dropping and the relative contribution of the thermal star
or disk could be rising.  By using, e.g., MATISSE \citep{Lopezetal2014} in combination with GRAVITY,
a centroid shift between images in, e.g., the $K$ and $J$ bands could be
detectable.  We explored this idea briefly in an earlier conference presentation
\citep{Markoff2008}, and found that for most sources this shift
would be a challenge to detect, however it is worth considering for
the future generation of interferometers.  The results would provide
powerful constraints of particle acceleration efficiency and cooling
in microquasar jets.

\section*{Acknowledgements}

Based on observations collected at the European Southern Observatory under ESO programme ID DDT 2101.D-0517. S.M. is thankful for support from an NWO (Netherlands Organisation for Scientific Research) VICI award, grant Nr. 639.043.513. J.D. was supported by a Sofja Kovalevskaja award from the Alexander von Humboldt foundation and in part by NSF grant AST 1909711. JCAM-J is the recipient of an Australian Research Council Future Fellowship (FT140101082), funded by the Australian government.




\bibliographystyle{mnras}
\bibliography{refs2019} 

\begin{thebibliography}{}
\makeatletter
\relax
\def\mn@urlcharsother{\let\do\@makeother \do\$\do\&\do\#\do\^\do\_\do\%\do\~}
\def\mn@doi{\begingroup\mn@urlcharsother \@ifnextchar [ {\mn@doi@}
  {\mn@doi@[]}}
\def\mn@doi@[#1]#2{\def\@tempa{#1}\ifx\@tempa\@empty \href
  {http://dx.doi.org/#2} {doi:#2}\else \href {http://dx.doi.org/#2} {#1}\fi
  \endgroup}
\def\mn@eprint#1#2{\mn@eprint@#1:#2::\@nil}
\def\mn@eprint@arXiv#1{\href {http://arxiv.org/abs/#1} {{\tt arXiv:#1}}}
\def\mn@eprint@dblp#1{\href {http://dblp.uni-trier.de/rec/bibtex/#1.xml}
  {dblp:#1}}
\def\mn@eprint@#1:#2:#3:#4\@nil{\def\@tempa {#1}\def\@tempb {#2}\def\@tempc
  {#3}\ifx \@tempc \@empty \let \@tempc \@tempb \let \@tempb \@tempa \fi \ifx
  \@tempb \@empty \def\@tempb {arXiv}\fi \@ifundefined
  {mn@eprint@\@tempb}{\@tempb:\@tempc}{\expandafter \expandafter \csname
  mn@eprint@\@tempb\endcsname \expandafter{\@tempc}}}

\bibitem[\protect\citeauthoryear{{Abeysekara} et~al.,}{{Abeysekara}
  et~al.}{2018}]{HAWCSS433_2018}
{Abeysekara} A.~U.,  et~al., 2018, \mn@doi [\nat] {10.1038/s41586-018-0565-5},
  \href {https://ui.adsabs.harvard.edu/abs/2018Natur.562...82A} {562, 82}

\bibitem[\protect\citeauthoryear{{Angel}, {Hill}, {Strittmatter}, {Salinari}
  \& {Weigelt}}{{Angel} et~al.}{1998}]{Angeletal1998}
{Angel} J. R.~P.,  {Hill} J.~M.,  {Strittmatter} P.~A.,  {Salinari} P.,
  {Weigelt} G.,  1998, in {Reasenberg} R.~D.,  ed.,  Society of Photo-Optical
  Instrumentation Engineers (SPIE) Conference Series Vol. 3350, \procspie. pp
  881--889, \mn@doi{10.1117/12.317156}

\bibitem[\protect\citeauthoryear{{Asada} \& {Nakamura}}{{Asada} \&
  {Nakamura}}{2012}]{AsadaNakamura2012}
{Asada} K.,  {Nakamura} M.,  2012, \mn@doi [\apjl]
  {10.1088/2041-8205/745/2/L28}, \href
  {http://cdsads.u-strasbg.fr/abs/2012ApJ...745L..28A} {745, L28}

\bibitem[\protect\citeauthoryear{{Atri} et~al.,}{{Atri}
  et~al.}{2019}]{Atrietal2019}
{Atri} P.,  et~al., 2019, arXiv e-prints, \href
  {https://ui.adsabs.harvard.edu/abs/2019arXiv190807199A} {p. arXiv:1908.07199}

\bibitem[\protect\citeauthoryear{{Atri} et~al.,}{{Atri}
  et~al.}{2020}]{Atri2020}
{Atri} P.,  et~al., 2020, \mn@doi [\mnras] {10.1093/mnrasl/slaa010}, \href
  {https://ui.adsabs.harvard.edu/abs/2020MNRAS.493L..81A} {493, L81}

\bibitem[\protect\citeauthoryear{{Baglio} et~al.,}{{Baglio}
  et~al.}{2018}]{Baglio2018}
{Baglio} M.~C.,  et~al., 2018, \mn@doi [\apj] {10.3847/1538-4357/aae532}, \href
  {https://ui.adsabs.harvard.edu/abs/2018ApJ...867..114B} {867, 114}

\bibitem[\protect\citeauthoryear{{Ball}, {Sironi}  \& {{\"O}zel}}{{Ball}
  et~al.}{2018}]{BallSironiOzel2018}
{Ball} D.,  {Sironi} L.,   {{\"O}zel} F.,  2018, \mn@doi [\apj]
  {10.3847/1538-4357/aac820}, \href
  {https://ui.adsabs.harvard.edu/abs/2018ApJ...862...80B} {862, 80}

\bibitem[\protect\citeauthoryear{{Bardeen} \& {Petterson}}{{Bardeen} \&
  {Petterson}}{1975}]{BardeenPetterson1975}
{Bardeen} J.~M.,  {Petterson} J.~A.,  1975, \mn@doi [ApJ] {10.1086/181711},
  \href {http://adsabs.harvard.edu/abs/1975ApJ...195L..65B} {195, L65}

\bibitem[\protect\citeauthoryear{{Belloni}}{{Belloni}}{2010}]{Belloni2010}
{Belloni} T.~M.,  2010, in {T.~Belloni} ed.,  Lecture Notes in Physics, Berlin
  Springer Verlag Vol. 794, Lecture Notes in Physics, Berlin Springer Verlag.
  p.~53 (\mn@eprint {arXiv} {0909.2474}), \mn@doi{10.1007/978-3-540-76937-8_3}

\bibitem[\protect\citeauthoryear{{Belloni}, {Homan}, {Casella}, {van der Klis},
  {Nespoli}, {Lewin}, {Miller}  \& {M{\'e}ndez}}{{Belloni}
  et~al.}{2005}]{Bellonietal2005}
{Belloni} T.,  {Homan} J.,  {Casella} P.,  {van der Klis} M.,  {Nespoli} E.,
  {Lewin} W.~H.~G.,  {Miller} J.~M.,   {M{\'e}ndez} M.,  2005, \mn@doi [\aap]
  {10.1051/0004-6361:20042457}, \href
  {https://ui.adsabs.harvard.edu/abs/2005A&A...440..207B} {440, 207}

\bibitem[\protect\citeauthoryear{{Bernardini}, {Russell}, {Kolojonen},
  {Stella}, {Hynes}  \& {Corbel}}{{Bernardini} et~al.}{2016}]{Bernardini2016}
{Bernardini} F.,  {Russell} D.~M.,  {Kolojonen} K.~I.~I.,  {Stella} L.,
  {Hynes} R.~I.,   {Corbel} S.,  2016, \mn@doi [\apj]
  {10.3847/0004-637X/826/2/149}, \href
  {https://ui.adsabs.harvard.edu/abs/2016ApJ...826..149B} {826, 149}

\bibitem[\protect\citeauthoryear{{Blundell}, {Schmidtobreick}  \&
  {Trushkin}}{{Blundell} et~al.}{2011}]{BlundellSchmidtobreickTrushkin2011}
{Blundell} K.~M.,  {Schmidtobreick} L.,   {Trushkin} S.,  2011, \mn@doi
  [\mnras] {10.1111/j.1365-2966.2011.18785.x}, \href
  {https://ui.adsabs.harvard.edu/abs/2011MNRAS.417.2401B} {417, 2401}

\bibitem[\protect\citeauthoryear{{Brocksopp}, {Corbel}, {Tzioumis},
  {Broderick}, {Rodriguez}, {Yang}, {Fender}  \& {Paragi}}{{Brocksopp}
  et~al.}{2013}]{Brocksoppetal2013}
{Brocksopp} C.,  {Corbel} S.,  {Tzioumis} A.,  {Broderick} J.~W.,  {Rodriguez}
  J.,  {Yang} J.,  {Fender} R.~P.,   {Paragi} Z.,  2013, \mn@doi [\mnras]
  {10.1093/mnras/stt493}, \href
  {https://ui.adsabs.harvard.edu/abs/2013MNRAS.432..931B} {432, 931}

\bibitem[\protect\citeauthoryear{{Buscher}, {Creech-Eakman}, {Farris}, {Haniff}
   \& {Young}}{{Buscher} et~al.}{2013}]{Buscher2013}
{Buscher} D.~F.,  {Creech-Eakman} M.,  {Farris} A.,  {Haniff} C.~A.,   {Young}
  J.~S.,  2013, \mn@doi [Journal of Astronomical Instrumentation]
  {10.1142/S2251171713400011}, \href
  {https://ui.adsabs.harvard.edu/abs/2013JAI.....240001B} {2, 1340001}

\bibitem[\protect\citeauthoryear{{Buxton} \& {Bailyn}}{{Buxton} \&
  {Bailyn}}{2004}]{Buxton2004}
{Buxton} M.~M.,  {Bailyn} C.~D.,  2004, \mn@doi [\apj] {10.1086/424503}, \href
  {https://ui.adsabs.harvard.edu/abs/2004ApJ...615..880B} {615, 880}

\bibitem[\protect\citeauthoryear{{Buxton}, {Bailyn}, {Capelo}, {Chatterjee},
  {Din{\c{c}}er}, {Kalemci}  \& {Tomsick}}{{Buxton}
  et~al.}{2012}]{Buxtonetal2012}
{Buxton} M.~M.,  {Bailyn} C.~D.,  {Capelo} H.~L.,  {Chatterjee} R.,
  {Din{\c{c}}er} T.,  {Kalemci} E.,   {Tomsick} J.~A.,  2012, \mn@doi [\aj]
  {10.1088/0004-6256/143/6/130}, \href
  {https://ui.adsabs.harvard.edu/abs/2012AJ....143..130B} {143, 130}

\bibitem[\protect\citeauthoryear{{Cantrell} et~al.,}{{Cantrell}
  et~al.}{2010}]{Cantrelletal2010}
{Cantrell} A.~G.,  et~al., 2010, \mn@doi [\apj] {10.1088/0004-637X/710/2/1127},
  \href {https://ui.adsabs.harvard.edu/abs/2010ApJ...710.1127C} {710, 1127}

\bibitem[\protect\citeauthoryear{{Casella} et~al.,}{{Casella}
  et~al.}{2010}]{Casellaetal2010}
{Casella} P.,  et~al., 2010, \mn@doi [\mnras]
  {10.1111/j.1745-3933.2010.00826.x}, \href
  {https://ui.adsabs.harvard.edu/abs/2010MNRAS.404L..21C} {404, L21}

\bibitem[\protect\citeauthoryear{{Chaty}, {Dubus}  \& {Raichoor}}{{Chaty}
  et~al.}{2011}]{Chaty2011}
{Chaty} S.,  {Dubus} G.,   {Raichoor} A.,  2011, \mn@doi [\aap]
  {10.1051/0004-6361/201015589}, \href
  {https://ui.adsabs.harvard.edu/abs/2011A&A...529A...3C} {529, A3}

\bibitem[\protect\citeauthoryear{{Che} et~al.,}{{Che}
  et~al.}{2011}]{Cheetal2011}
{Che} X.,  et~al., 2011, \mn@doi [ApJ] {10.1088/0004-637X/732/2/68}, \href
  {http://cdsads.u-strasbg.fr/abs/2011ApJ...732...68C} {732, 68}

\bibitem[\protect\citeauthoryear{{Cherepashchuk}, {Katysheva}, {Khruzina},
  {Shugarov}, {Tatarnikov}, {Burlak}  \& {Shatsky}}{{Cherepashchuk}
  et~al.}{2019}]{Cherepashchuk2019}
{Cherepashchuk} A.~M.,  {Katysheva} N.~A.,  {Khruzina} T.~S.,  {Shugarov}
  S.~Y.,  {Tatarnikov} A.~M.,  {Burlak} M.~A.,   {Shatsky} N.~I.,  2019,
  \mn@doi [\mnras] {10.1093/mnras/sty3166}, \href
  {https://ui.adsabs.harvard.edu/abs/2019MNRAS.483.1067C} {483, 1067}

\bibitem[\protect\citeauthoryear{{Chevalier}, {Ilovaisky}, {van Paradijs},
  {Pedersen}  \& {van der Klis}}{{Chevalier} et~al.}{1989}]{Chevalieretal1989}
{Chevalier} C.,  {Ilovaisky} S.~A.,  {van Paradijs} J.,  {Pedersen} H.,   {van
  der Klis} M.,  1989, \aap, \href
  {https://ui.adsabs.harvard.edu/abs/1989A&A...210..114C} {210, 114}

\bibitem[\protect\citeauthoryear{{Clarkson}, {Charles}  \& {Onyett}}{{Clarkson}
  et~al.}{2004}]{Clarksonetal2004}
{Clarkson} W.~I.,  {Charles} P.~A.,   {Onyett} N.,  2004, \mn@doi [\mnras]
  {10.1111/j.1365-2966.2004.07293.x}, \href
  {https://ui.adsabs.harvard.edu/abs/2004MNRAS.348..458C} {348, 458}

\bibitem[\protect\citeauthoryear{{Connors} et~al.,}{{Connors}
  et~al.}{2017}]{Connorsetal2017}
{Connors} R.~M.~T.,  et~al., 2017, \mn@doi [\mnras] {10.1093/mnras/stw3150},
  \href {https://ui.adsabs.harvard.edu/abs/2017MNRAS.466.4121C} {466, 4121}

\bibitem[\protect\citeauthoryear{Corbel \& Fender}{Corbel \&
  Fender}{2002}]{CorbelFender2002}
Corbel S.,  Fender R.,  2002, ApJ, 573, L35

\bibitem[\protect\citeauthoryear{{Corbel}, {Fender}, {Tzioumis}, {Tomsick},
  {Orosz}, {Miller}, {Wijnands}  \& {Kaaret}}{{Corbel}
  et~al.}{2002}]{Corbeletal2002}
{Corbel} S.,  {Fender} R.~P.,  {Tzioumis} A.~K.,  {Tomsick} J.~A.,  {Orosz}
  J.~A.,  {Miller} J.~M.,  {Wijnands} R.,   {Kaaret} P.,  2002, Science, \href
  {http://adsabs.harvard.edu/cgi-bin/nph-bib_query?bibcode=2002Sci...298..196C&amp;db_key=AST}
  {298, 196}

\bibitem[\protect\citeauthoryear{{Corbel} et~al.,}{{Corbel}
  et~al.}{2012}]{Corbeletal2012}
{Corbel} S.,  et~al., 2012, \mn@doi [MNRAS] {10.1111/j.1365-2966.2012.20517.x},
  \href {http://cdsads.u-strasbg.fr/abs/2012MNRAS.421.2947C} {421, 2947}

\bibitem[\protect\citeauthoryear{{Corbel}, {Coriat}, {Brocksopp}, {Tzioumis},
  {Fender}, {Tomsick}, {Buxton}  \& {Bailyn}}{{Corbel}
  et~al.}{2013}]{Corbeletal2013}
{Corbel} S.,  {Coriat} M.,  {Brocksopp} C.,  {Tzioumis} A.~K.,  {Fender} R.~P.,
   {Tomsick} J.~A.,  {Buxton} M.~M.,   {Bailyn} C.~D.,  2013, \mn@doi [MNRAS]
  {10.1093/mnras/sts215}, \href
  {http://cdsads.u-strasbg.fr/abs/2013MNRAS.428.2500C} {428, 2500}

\bibitem[\protect\citeauthoryear{{Corbet}}{{Corbet}}{2003}]{Corbet2003}
{Corbet} R. H.~D.,  2003, \mn@doi [\apj] {10.1086/377469}, \href
  {https://ui.adsabs.harvard.edu/abs/2003ApJ...595.1086C} {595, 1086}

\bibitem[\protect\citeauthoryear{{Corbet}, {Pearlman}, {Buxton}  \&
  {Levine}}{{Corbet} et~al.}{2010}]{Corbetetal2010}
{Corbet} R. H.~D.,  {Pearlman} A.~B.,  {Buxton} M.,   {Levine} A.~M.,  2010,
  \mn@doi [\apj] {10.1088/0004-637X/719/1/979}, \href
  {https://ui.adsabs.harvard.edu/abs/2010ApJ...719..979C} {719, 979}

\bibitem[\protect\citeauthoryear{{Denisenko}}{{Denisenko}}{2018}]{Denisenko2018}
{Denisenko} D.,  2018, The Astronomer's Telegram, \href
  {https://ui.adsabs.harvard.edu/abs/2018ATel11400....1D} {11400, 1}

\bibitem[\protect\citeauthoryear{{Dhawan}, {Mirabel}  \&
  {Rodr{\'\i}guez}}{{Dhawan} et~al.}{2000}]{DhawanMirabelRodriguez2000}
{Dhawan} V.,  {Mirabel} I.~F.,   {Rodr{\'\i}guez} L.~F.,  2000, \mn@doi [\apj]
  {10.1086/317088}, \href
  {https://ui.adsabs.harvard.edu/abs/2000ApJ...543..373D} {543, 373}

\bibitem[\protect\citeauthoryear{{Doroshenko}, {Santangelo}, {Suleimanov},
  {Kreykenbohm}, {Staubert}, {Ferrigno}  \& {Klochkov}}{{Doroshenko}
  et~al.}{2010}]{Doroshenkoetal2010}
{Doroshenko} V.,  {Santangelo} A.,  {Suleimanov} V.,  {Kreykenbohm} I.,
  {Staubert} R.,  {Ferrigno} C.,   {Klochkov} D.,  2010, \mn@doi [\aap]
  {10.1051/0004-6361/200912951}, \href
  {https://ui.adsabs.harvard.edu/abs/2010A&A...515A..10D} {515, A10}

\bibitem[\protect\citeauthoryear{{Eikenberry}, {Matthews}, {Morgan},
  {Remillard}  \& {Nelson}}{{Eikenberry} et~al.}{1998}]{Eikenberryetal1998}
{Eikenberry} S.~S.,  {Matthews} K.,  {Morgan} E.~H.,  {Remillard} R.~A.,
  {Nelson} R.~W.,  1998, ApJ, 494, L61

\bibitem[\protect\citeauthoryear{{Espinasse} et~al.,}{{Espinasse}
  et~al.}{2020}]{Espinasseetal2020}
{Espinasse} M.,  et~al., 2020, \apjl, \href
  {https://ui.adsabs.harvard.edu/abs/2020arXiv200406416E} {pp "in press,
  arXiv:2004.06416"}

\bibitem[\protect\citeauthoryear{{Event Horizon Telescope Collaboration}
  et~al.,}{{Event Horizon Telescope Collaboration} et~al.}{2019}]{EHT2019a}
{Event Horizon Telescope Collaboration} et~al., 2019, \mn@doi [\apjl]
  {10.3847/2041-8213/ab0ec7}, \href
  {https://ui.adsabs.harvard.edu/abs/2019ApJ...875L...1E} {875, L1}

\bibitem[\protect\citeauthoryear{Falcke, K\"ording  \& Markoff}{Falcke
  et~al.}{2004}]{FalckeKoerdingMarkoff2004}
Falcke H.,  K\"ording E.,   Markoff S.,  2004, \aap, 414, 895

\bibitem[\protect\citeauthoryear{{Fender}, {Pooley}, {Brocksopp}  \&
  {Newell}}{{Fender} et~al.}{1997}]{Fenderetal1997}
{Fender} R.~P.,  {Pooley} G.~G.,  {Brocksopp} C.,   {Newell} S.~J.,  1997,
  MNRAS, 290, L65

\bibitem[\protect\citeauthoryear{{Fender}, {Garrington}, {McKay}, {Muxlow},
  {Pooley}, {Spencer}, {Stirling}  \& {Waltman}}{{Fender}
  et~al.}{1999}]{Fenderetal1999}
{Fender} R.~P.,  {Garrington} S.~T.,  {McKay} D.~J.,  {Muxlow} T.~W.~B.,
  {Pooley} G.~G.,  {Spencer} R.~E.,  {Stirling} A.~M.,   {Waltman} E.~B.,
  1999, MNRAS, 304, 865

\bibitem[\protect\citeauthoryear{{Fender}, {Pooley}, {Durouchoux}, {Tilanus}
  \& {Brocksopp}}{{Fender} et~al.}{2000}]{Fenderetal2000}
{Fender} R.~P.,  {Pooley} G.~G.,  {Durouchoux} P.,  {Tilanus} R.~P.~J.,
  {Brocksopp} C.,  2000, \mn@doi [\mnras] {10.1046/j.1365-8711.2000.03219.x},
  \href {https://ui.adsabs.harvard.edu/abs/2000MNRAS.312..853F} {312, 853}

\bibitem[\protect\citeauthoryear{{Fender}, {Homan}  \& {Belloni}}{{Fender}
  et~al.}{2009}]{FenderHomanBelloni2009}
{Fender} R.~P.,  {Homan} J.,   {Belloni} T.~M.,  2009, \mn@doi [MNRAS]
  {10.1111/j.1365-2966.2009.14841.x}, \href
  {http://cdsads.u-strasbg.fr/abs/2009MNRAS.396.1370F} {396, 1370}

\bibitem[\protect\citeauthoryear{{Fender} et~al.,}{{Fender}
  et~al.}{2017}]{Fenderetal2017}
{Fender} R.,  et~al., 2017, arXiv e-prints, \href
  {https://ui.adsabs.harvard.edu/abs/2017arXiv171104132F} {p. arXiv:1711.04132}

\bibitem[\protect\citeauthoryear{{Gallo}, {Fender}, {Kaiser}, {Russell},
  {Morganti}, {Oosterloo}  \& {Heinz}}{{Gallo} et~al.}{2005}]{Galloetal2005}
{Gallo} E.,  {Fender} R.,  {Kaiser} C.,  {Russell} D.,  {Morganti} R.,
  {Oosterloo} T.,   {Heinz} S.,  2005, \mn@doi [Nature] {10.1038/nature03879},
  \href {http://adsabs.harvard.edu/abs/2005Natur.436..819G} {436, 819}

\bibitem[\protect\citeauthoryear{{Gandhi} et~al.,}{{Gandhi}
  et~al.}{2011}]{Gandhietal2011}
{Gandhi} P.,  et~al., 2011, \mn@doi [ApJ] {10.1088/2041-8205/740/1/L13}, \href
  {http://cdsads.u-strasbg.fr/abs/2011ApJ...740L..13G} {740, L13}

\bibitem[\protect\citeauthoryear{{Gandhi} et~al.,}{{Gandhi}
  et~al.}{2017}]{Gandhi2017NatAs}
{Gandhi} P.,  et~al., 2017, \mn@doi [Nature Astronomy]
  {10.1038/s41550-017-0273-3}, \href
  {https://ui.adsabs.harvard.edu/abs/2017NatAs...1..859G} {1, 859}

\bibitem[\protect\citeauthoryear{{Gandhi}, {Rao}, {Johnson}, {Paice}  \&
  {Maccarone}}{{Gandhi} et~al.}{2019}]{Gandhi2019GAIA}
{Gandhi} P.,  {Rao} A.,  {Johnson} M. A.~C.,  {Paice} J.~A.,   {Maccarone}
  T.~J.,  2019, \mn@doi [\mnras] {10.1093/mnras/stz438}, \href
  {https://ui.adsabs.harvard.edu/abs/2019MNRAS.485.2642G} {485, 2642}

\bibitem[\protect\citeauthoryear{{Gies}, {ten Brummelaar}, {Schaefer}, {Baron}
  \& {White}}{{Gies} et~al.}{2019}]{Gies2019}
{Gies} D.,  {ten Brummelaar} T.,  {Schaefer} G.,  {Baron} F.,   {White} R.,
  2019, in \baas. p.~226

\bibitem[\protect\citeauthoryear{{Gonz{\'a}lez Hern{\'a}ndez} \&
  {Casares}}{{Gonz{\'a}lez Hern{\'a}ndez} \&
  {Casares}}{2010}]{GonzalezHernandezCasares2010}
{Gonz{\'a}lez Hern{\'a}ndez} J.~I.,  {Casares} J.,  2010, \mn@doi [\aap]
  {10.1051/0004-6361/201014088}, \href
  {https://ui.adsabs.harvard.edu/abs/2010A&A...516A..58G} {516, A58}

\bibitem[\protect\citeauthoryear{{Gravity Collaboration} et~al.,}{{Gravity
  Collaboration} et~al.}{2017a}]{GRAVITY2017}
{Gravity Collaboration} et~al., 2017a, \mn@doi [\aap]
  {10.1051/0004-6361/201730838}, \href
  {https://ui.adsabs.harvard.edu/abs/2017A%26A...602A..94G} {602, A94}

\bibitem[\protect\citeauthoryear{{Gravity Collaboration} et~al.,}{{Gravity
  Collaboration} et~al.}{2017b}]{GRAVITY2017SS433}
{Gravity Collaboration} et~al., 2017b, \mn@doi [\aap]
  {10.1051/0004-6361/201731038}, \href
  {https://ui.adsabs.harvard.edu/abs/2017A&A...602L..11G} {602, L11}

\bibitem[\protect\citeauthoryear{{Gravity Collaboration} et~al.,}{{Gravity
  Collaboration} et~al.}{2018a}]{GRAVITY2018quasar}
{Gravity Collaboration} et~al., 2018a, \mn@doi [\nat]
  {10.1038/s41586-018-0731-9}, \href
  {https://ui.adsabs.harvard.edu/abs/2018Natur.563..657G} {563, 657}

\bibitem[\protect\citeauthoryear{{Gravity Collaboration} et~al.,}{{Gravity
  Collaboration} et~al.}{2018b}]{GRAVITY2018S2}
{Gravity Collaboration} et~al., 2018b, \mn@doi [\aap]
  {10.1051/0004-6361/201833718}, \href
  {https://ui.adsabs.harvard.edu/abs/2018A&A...615L..15G} {615, L15}

\bibitem[\protect\citeauthoryear{{Gravity Collaboration} et~al.,}{{Gravity
  Collaboration} et~al.}{2018c}]{GRAVITY2018sgra}
{Gravity Collaboration} et~al., 2018c, \mn@doi [\aap]
  {10.1051/0004-6361/201834294}, \href
  {https://ui.adsabs.harvard.edu/abs/2018A&A...618L..10G} {618, L10}

\bibitem[\protect\citeauthoryear{{Gravity Collaboration} et~al.,}{{Gravity
  Collaboration} et~al.}{2019a}]{GRAVITY2019exoplanet}
{Gravity Collaboration} et~al., 2019a, \mn@doi [\aap]
  {10.1051/0004-6361/201935253}, \href
  {https://ui.adsabs.harvard.edu/abs/2019A&A...623L..11G} {623, L11}

\bibitem[\protect\citeauthoryear{{Gravity Collaboration} et~al.,}{{Gravity
  Collaboration} et~al.}{2019b}]{GRAVITY2019S2}
{Gravity Collaboration} et~al., 2019b, \mn@doi [\aap]
  {10.1051/0004-6361/201935656}, \href
  {https://ui.adsabs.harvard.edu/abs/2019A&A...625L..10G} {625, L10}

\bibitem[\protect\citeauthoryear{{Gravity Collaboration} et~al.,}{{Gravity
  Collaboration} et~al.}{2020}]{GRAVITY2020precession}
{Gravity Collaboration} et~al., 2020, arXiv e-prints, \href
  {https://ui.adsabs.harvard.edu/abs/2020arXiv200407187G} {pp "A\&A Lett. in
  press, arXiv:2004.07187"}

\bibitem[\protect\citeauthoryear{{Greene}, {Bailyn}  \& {Orosz}}{{Greene}
  et~al.}{2001}]{Greeneetal2001}
{Greene} J.,  {Bailyn} C.~D.,   {Orosz} J.~A.,  2001, \mn@doi [\apj]
  {10.1086/321411}, \href
  {https://ui.adsabs.harvard.edu/abs/2001ApJ...554.1290G} {554, 1290}

\bibitem[\protect\citeauthoryear{{Hammerstein}, {Cackett}, {Reynolds}  \&
  {Miller}}{{Hammerstein} et~al.}{2018}]{Hammersteinetal2018}
{Hammerstein} E.~K.,  {Cackett} E.~M.,  {Reynolds} M.~T.,   {Miller} J.~M.,
  2018, \mn@doi [\mnras] {10.1093/mnras/sty1348}, \href
  {https://ui.adsabs.harvard.edu/abs/2018MNRAS.478.4317H} {478, 4317}

\bibitem[\protect\citeauthoryear{{Hillwig} \& {Gies}}{{Hillwig} \&
  {Gies}}{2008}]{Hillwig2008}
{Hillwig} T.~C.,  {Gies} D.~R.,  2008, \mn@doi [\apjl] {10.1086/587140}, \href
  {https://ui.adsabs.harvard.edu/abs/2008ApJ...676L..37H} {676, L37}

\bibitem[\protect\citeauthoryear{{Hjellming} \& {Rupen}}{{Hjellming} \&
  {Rupen}}{1995}]{HjellmingRupen1995}
{Hjellming} R.~M.,  {Rupen} M.~P.,  1995, Nature, 375, 464

\bibitem[\protect\citeauthoryear{{Hjellming} et~al.,}{{Hjellming}
  et~al.}{2000}]{Hjellmingetal2000}
{Hjellming} R.~M.,  et~al., 2000, ApJ, \href
  {http://adsabs.harvard.edu/cgi-bin/nph-bib_query?bibcode=2000ApJ...544..977H&db_key=AST}
  {544, 977}

\bibitem[\protect\citeauthoryear{{Homan}, {Buxton}, {Markoff}, {Bailyn},
  {Nespoli}  \& {Belloni}}{{Homan} et~al.}{2005}]{Homanetal2005}
{Homan} J.,  {Buxton} M.,  {Markoff} S.,  {Bailyn} C.~D.,  {Nespoli} E.,
  {Belloni} T.,  2005, \mn@doi [ApJ] {10.1086/428722}, \href
  {http://adsabs.harvard.edu/cgi-bin/nph-bib_query?bibcode=2005ApJ...624..295H&db_key=AST}
  {624, 295}

\bibitem[\protect\citeauthoryear{{Hynes}}{{Hynes}}{2005}]{Hynes2005}
{Hynes} R.~I.,  2005, \mn@doi [ApJ] {10.1086/428445}, \href
  {http://adsabs.harvard.edu/abs/2005ApJ...623.1026H} {623, 1026}

\bibitem[\protect\citeauthoryear{{Jonker} \& {Nelemans}}{{Jonker} \&
  {Nelemans}}{2004}]{JonkerNelemans2004}
{Jonker} P.~G.,  {Nelemans} G.,  2004, \mn@doi [\mnras]
  {10.1111/j.1365-2966.2004.08193.x}, \href
  {https://ui.adsabs.harvard.edu/abs/2004MNRAS.354..355J} {354, 355}

\bibitem[\protect\citeauthoryear{{Jonker}, {Nelemans}  \& {Bassa}}{{Jonker}
  et~al.}{2007}]{Jonkeretal2007}
{Jonker} P.~G.,  {Nelemans} G.,   {Bassa} C.~G.,  2007, \mn@doi [\mnras]
  {10.1111/j.1365-2966.2006.11210.x}, \href
  {https://ui.adsabs.harvard.edu/abs/2007MNRAS.374..999J} {374, 999}

\bibitem[\protect\citeauthoryear{{Kalamkar}, {Casella}, {Uttley}, {O'Brien},
  {Russell}, {Maccarone}, {van der Klis}  \& {Vincentelli}}{{Kalamkar}
  et~al.}{2016}]{Kalamkar2016}
{Kalamkar} M.,  {Casella} P.,  {Uttley} P.,  {O'Brien} K.,  {Russell} D.,
  {Maccarone} T.,  {van der Klis} M.,   {Vincentelli} F.,  2016, \mn@doi
  [\mnras] {10.1093/mnras/stw1211}, \href
  {https://ui.adsabs.harvard.edu/abs/2016MNRAS.460.3284K} {460, 3284}

\bibitem[\protect\citeauthoryear{{Kaper}, {Lamers}, {Ruymaekers}, {van den
  Heuvel}  \& {Zuiderwijk}}{{Kaper} et~al.}{1995}]{Kaper1995}
{Kaper} L.,  {Lamers} H.~J.~G.~L.~M.,  {Ruymaekers} E.,  {van den Heuvel}
  E.~P.~J.,   {Zuiderwijk} E.~J.,  1995, \aap, \href
  {https://ui.adsabs.harvard.edu/abs/1995A&A...300..446K} {300, 446}

\bibitem[\protect\citeauthoryear{{Kawamuro} et~al.,}{{Kawamuro}
  et~al.}{2018}]{Kawamuro2018}
{Kawamuro} T.,  et~al., 2018, The Astronomer's Telegram, \href
  {https://ui.adsabs.harvard.edu/abs/2018ATel11399....1K} {11399, 1}

\bibitem[\protect\citeauthoryear{{Khargharia}, {Froning}  \&
  {Robinson}}{{Khargharia} et~al.}{2010}]{KharghariaFroningRobinson2010}
{Khargharia} J.,  {Froning} C.~S.,   {Robinson} E.~L.,  2010, \mn@doi [\apj]
  {10.1088/0004-637X/716/2/1105}, \href
  {https://ui.adsabs.harvard.edu/abs/2010ApJ...716.1105K} {716, 1105}

\bibitem[\protect\citeauthoryear{{Kishimoto}, {H{\"o}nig}, {Antonucci},
  {Barvainis}, {Kotani}, {Tristram}, {Weigelt}  \& {Levin}}{{Kishimoto}
  et~al.}{2011}]{kishimoto2011}
{Kishimoto} M.,  {H{\"o}nig} S.~F.,  {Antonucci} R.,  {Barvainis} R.,  {Kotani}
  T.,  {Tristram} K.~R.~W.,  {Weigelt} G.,   {Levin} K.,  2011, \mn@doi [\aap]
  {10.1051/0004-6361/201016054}, \href
  {https://ui.adsabs.harvard.edu/abs/2011A&A...527A.121K} {527, A121}

\bibitem[\protect\citeauthoryear{{Koljonen} et~al.,}{{Koljonen}
  et~al.}{2015}]{Koljonenetal2015}
{Koljonen} K.~I.~I.,  et~al., 2015, \mn@doi [\apj]
  {10.1088/0004-637X/814/2/139}, \href
  {http://cdsads.u-strasbg.fr/abs/2015ApJ...814..139K} {814, 139}

\bibitem[\protect\citeauthoryear{{Lachaume}}{{Lachaume}}{2003}]{lachaume2003}
{Lachaume} R.,  2003, \mn@doi [\aap] {10.1051/0004-6361:20030072}, \href
  {https://ui.adsabs.harvard.edu/abs/2003A&A...400..795L} {400, 795}

\bibitem[\protect\citeauthoryear{{Lacour} et~al.,}{{Lacour}
  et~al.}{2019}]{lacour2019}
{Lacour} S.,  et~al., 2019, \mn@doi [\aap] {10.1051/0004-6361/201834981}, \href
  {https://ui.adsabs.harvard.edu/abs/2019A&A...624A..99L} {624, A99}

\bibitem[\protect\citeauthoryear{{Lapeyrere} et~al.,}{{Lapeyrere}
  et~al.}{2014}]{lapeyrere2014}
{Lapeyrere} V.,  et~al., 2014, in \procspie. p. 91462D,
  \mn@doi{10.1117/12.2056850}

\bibitem[\protect\citeauthoryear{{Laurent}, {Rodriguez}, {Wilms}, {Cadolle
  Bel}, {Pottschmidt}  \& {Grinberg}}{{Laurent} et~al.}{2011}]{Laurentetal2011}
{Laurent} P.,  {Rodriguez} J.,  {Wilms} J.,  {Cadolle Bel} M.,  {Pottschmidt}
  K.,   {Grinberg} V.,  2011, \mn@doi [Science] {10.1126/science.1200848},
  \href {http://cdsads.u-strasbg.fr/abs/2011Sci...332..438L} {332, 438}

\bibitem[\protect\citeauthoryear{{Liska}, {Hesp}, {Tchekhovskoy}, {Ingram},
  {van der Klis}  \& {Markoff}}{{Liska} et~al.}{2018}]{Liskaetal2018}
{Liska} M.,  {Hesp} C.,  {Tchekhovskoy} A.,  {Ingram} A.,  {van der Klis} M.,
  {Markoff} S.,  2018, \mn@doi [\mnras] {10.1093/mnrasl/slx174}, \href
  {https://ui.adsabs.harvard.edu/abs/2018MNRAS.474L..81L} {474, L81}

\bibitem[\protect\citeauthoryear{{Lopez}, {Marshall}, {Canizares}, {Schulz}  \&
  {Kane}}{{Lopez} et~al.}{2006}]{Lopezetal2006}
{Lopez} L.~A.,  {Marshall} H.~L.,  {Canizares} C.~R.,  {Schulz} N.~S.,   {Kane}
  J.~F.,  2006, \mn@doi [\apj] {10.1086/506174}, \href
  {https://ui.adsabs.harvard.edu/abs/2006ApJ...650..338L} {650, 338}

\bibitem[\protect\citeauthoryear{{Lopez} et~al.,}{{Lopez}
  et~al.}{2014}]{Lopezetal2014}
{Lopez} B.,  et~al., 2014, The Messenger, \href
  {https://ui.adsabs.harvard.edu/abs/2014Msngr.157....5L} {157, 5}

\bibitem[\protect\citeauthoryear{{MacDonald} et~al.,}{{MacDonald}
  et~al.}{2014}]{MacDonaldetal2014}
{MacDonald} R. K.~D.,  et~al., 2014, \mn@doi [\apj]
  {10.1088/0004-637X/784/1/2}, \href
  {https://ui.adsabs.harvard.edu/abs/2014ApJ...784....2M} {784, 2}

\bibitem[\protect\citeauthoryear{{Maccarone}}{{Maccarone}}{2002}]{Maccarone2002}
{Maccarone} T.~J.,  2002, MNRAS, \href
  {http://adsabs.harvard.edu/cgi-bin/nph-bib_query?bibcode=2002MNRAS.336.1371M&db_key=AST}
  {336, 1371}

\bibitem[\protect\citeauthoryear{{Maitra}, {Scarpaci}, {Grinberg}, {Reynolds},
  {Markoff}, {Maccarone}  \& {Hynes}}{{Maitra} et~al.}{2017}]{Maitraetal2017}
{Maitra} D.,  {Scarpaci} J.~F.,  {Grinberg} V.,  {Reynolds} M.~T.,  {Markoff}
  S.,  {Maccarone} T.~J.,   {Hynes} R.~I.,  2017, \mn@doi [\apj]
  {10.3847/1538-4357/aa98a0}, \href
  {https://ui.adsabs.harvard.edu/abs/2017ApJ...851..148M} {851, 148}

\bibitem[\protect\citeauthoryear{{Malzac} et~al.,}{{Malzac}
  et~al.}{2018}]{Malzac2018}
{Malzac} J.,  et~al., 2018, \mn@doi [\mnras] {10.1093/mnras/sty2006}, \href
  {https://ui.adsabs.harvard.edu/abs/2018MNRAS.480.2054M} {480, 2054}

\bibitem[\protect\citeauthoryear{{Mandal}, {Singh}, {Stalin}, {Chandra}  \&
  {Gandhi}}{{Mandal} et~al.}{2018}]{Mandal2018}
{Mandal} A.~K.,  {Singh} A.,  {Stalin} C.~S.,  {Chandra} S.,   {Gandhi} P.,
  2018, The Astronomer's Telegram, \href
  {https://ui.adsabs.harvard.edu/abs/2018ATel11462....1M} {11462, 1}

\bibitem[\protect\citeauthoryear{{Markoff}}{{Markoff}}{2008}]{Markoff2008}
{Markoff} S.,  2008, in Microquasars and Beyond.  (\mn@eprint {arXiv}
  {0811.3601})

\bibitem[\protect\citeauthoryear{{Markoff}, {Falcke}  \& {Fender}}{{Markoff}
  et~al.}{2001}]{MarkoffFalckeFender2001}
{Markoff} S.,  {Falcke} H.,   {Fender} R.,  2001, \aap, 372, L25

\bibitem[\protect\citeauthoryear{{Markoff}, {Nowak}  \& {Wilms}}{{Markoff}
  et~al.}{2005}]{MarkoffNowakWilms2005}
{Markoff} S.,  {Nowak} M.~A.,   {Wilms} J.,  2005, \mn@doi [ApJ]
  {10.1086/497628}, \href {http://adsabs.harvard.edu/abs/2005ApJ...635.1203M}
  {635, 1203}

\bibitem[\protect\citeauthoryear{{Masetti}, {Bianchini}, {Bonibaker}, {della
  Valle}  \& {Vio}}{{Masetti} et~al.}{1996}]{Masettietal1996}
{Masetti} N.,  {Bianchini} A.,  {Bonibaker} J.,  {della Valle} M.,   {Vio} R.,
  1996, \aap, \href {https://ui.adsabs.harvard.edu/abs/1996A&A...314..123M}
  {314, 123}

\bibitem[\protect\citeauthoryear{{McClintock} \& {Remillard}}{{McClintock} \&
  {Remillard}}{2006}]{McClintockRemillard2006}
{McClintock} J.~E.,  {Remillard} R.~A.,  2006, {Black hole binaries}.
Compact stellar X-ray sources, Eds. Walter Lewin \& Michiel van der Klis.
  Cambridge Astrophysics Series, No. 39. Cambridge, UK: Cambridge University
  Press, pp 157--213

\bibitem[\protect\citeauthoryear{{McHardy}, {Koerding}, {Knigge}, {Uttley}  \&
  {Fender}}{{McHardy} et~al.}{2006}]{McHardyetal2006}
{McHardy} I.~M.,  {Koerding} E.,  {Knigge} C.,  {Uttley} P.,   {Fender} R.~P.,
  2006, \mn@doi [Nature] {10.1038/nature05389}, \href
  {http://adsabs.harvard.edu/cgi-bin/nph-bib_query?bibcode=2006Natur.444..730M&db_key=AST}
  {444, 730}

\bibitem[\protect\citeauthoryear{{Merloni}, {Heinz}  \& {di Matteo}}{{Merloni}
  et~al.}{2003}]{MerloniHeinzDiMatteo2003}
{Merloni} A.,  {Heinz} S.,   {di Matteo} T.,  2003, MNRAS, \href
  {http://adsabs.harvard.edu/cgi-bin/nph-bib_query?bibcode=2003MNRAS.345.1057M&amp;db_key=AST}
  {345, 1057}

\bibitem[\protect\citeauthoryear{{Merloni} et~al.,}{{Merloni}
  et~al.}{2012}]{Merlonietal2012}
{Merloni} A.,  et~al., 2012, arXiv e-prints, \href
  {https://ui.adsabs.harvard.edu/abs/2012arXiv1209.3114M} {p. arXiv:1209.3114}

\bibitem[\protect\citeauthoryear{{Migliari}, {Tomsick}, {Maccarone}, {Gallo},
  {Fender}, {Nelemans}  \& {Russell}}{{Migliari}
  et~al.}{2006}]{Migliarietal2006}
{Migliari} S.,  {Tomsick} J.~A.,  {Maccarone} T.~J.,  {Gallo} E.,  {Fender}
  R.~P.,  {Nelemans} G.,   {Russell} D.~M.,  2006, \mn@doi [ApJ]
  {10.1086/505028}, \href {http://cdsads.u-strasbg.fr/abs/2006ApJ...643L..41M}
  {643, L41}

\bibitem[\protect\citeauthoryear{{Migliari} et~al.,}{{Migliari}
  et~al.}{2007}]{Migliarietal2007}
{Migliari} S.,  et~al., 2007, \mn@doi [ApJ] {10.1086/522023}, \href
  {http://cdsads.u-strasbg.fr/abs/2007ApJ...670..610M} {670, 610}

\bibitem[\protect\citeauthoryear{{Miller-Jones}}{{Miller-Jones}}{2014}]{MillerJones2014}
{Miller-Jones} J. C.~A.,  2014, \mn@doi [\pasa] {10.1017/pasa.2014.7}, \href
  {https://ui.adsabs.harvard.edu/abs/2014PASA...31...16M} {31, e016}

\bibitem[\protect\citeauthoryear{{Miller-Jones}, {Fender}  \&
  {Nakar}}{{Miller-Jones} et~al.}{2006}]{Miller-Jones2006}
{Miller-Jones} J.~C.~A.,  {Fender} R.~P.,   {Nakar} E.,  2006, \mn@doi [\mnras]
  {10.1111/j.1365-2966.2006.10092.x}, \href
  {https://ui.adsabs.harvard.edu/abs/2006MNRAS.367.1432M} {367, 1432}

\bibitem[\protect\citeauthoryear{{Miller-Jones}, {Jonker}, {Ratti}, {Torres},
  {Brocksopp}, {Yang}  \& {Morrell}}{{Miller-Jones}
  et~al.}{2011}]{Miller-Jones2011}
{Miller-Jones} J.~C.~A.,  {Jonker} P.~G.,  {Ratti} E.~M.,  {Torres} M.~A.~P.,
  {Brocksopp} C.,  {Yang} J.,   {Morrell} N.~I.,  2011, \mn@doi [\mnras]
  {10.1111/j.1365-2966.2011.18704.x}, \href
  {https://ui.adsabs.harvard.edu/abs/2011MNRAS.415..306M} {415, 306}

\bibitem[\protect\citeauthoryear{{Miller-Jones} et~al.,}{{Miller-Jones}
  et~al.}{2012}]{Miller-Jonesetal2012}
{Miller-Jones} J.~C.~A.,  et~al., 2012, \mn@doi [MNRAS]
  {10.1111/j.1365-2966.2011.20326.x}, \href
  {http://cdsads.u-strasbg.fr/abs/2012MNRAS.421..468M} {421, 468}

\bibitem[\protect\citeauthoryear{{Miller-Jones} et~al.,}{{Miller-Jones}
  et~al.}{2019}]{MillerJonesetal2019Nature}
{Miller-Jones} J. C.~A.,  et~al., 2019, \mn@doi [\nat]
  {10.1038/s41586-019-1152-0}, \href
  {https://ui.adsabs.harvard.edu/abs/2019Natur.569..374M} {569, 374}

\bibitem[\protect\citeauthoryear{{Minniti} et~al.,}{{Minniti}
  et~al.}{2010}]{Minnitietal2010}
{Minniti} D.,  et~al., 2010, \mn@doi [\na] {10.1016/j.newast.2009.12.002},
  \href {https://ui.adsabs.harvard.edu/abs/2010NewA...15..433M} {15, 433}

\bibitem[\protect\citeauthoryear{{Mioduszewski}, {Rupen}, {Hjellming}, {Pooley}
   \& {Waltman}}{{Mioduszewski} et~al.}{2001}]{Mioduszewskietal2001}
{Mioduszewski} A.~J.,  {Rupen} M.~P.,  {Hjellming} R.~M.,  {Pooley} G.~G.,
  {Waltman} E.~B.,  2001, \mn@doi [\apj] {10.1086/320965}, \href
  {https://ui.adsabs.harvard.edu/abs/2001ApJ...553..766M} {553, 766}

\bibitem[\protect\citeauthoryear{{Mirabel} \& {Rodr{\'i}guez}}{{Mirabel} \&
  {Rodr{\'i}guez}}{1994}]{MirabelRodriguez1994}
{Mirabel} I.~F.,  {Rodr{\'i}guez} L.~F.,  1994, Nature, 371, 46

\bibitem[\protect\citeauthoryear{{Mirabel}, {Dhawan}, {Chaty}, {Rodriguez},
  {Marti}, {Robinson}, {Swank}  \& {Geballe}}{{Mirabel}
  et~al.}{1998}]{Mirabeletal1998}
{Mirabel} I.~F.,  {Dhawan} V.,  {Chaty} S.,  {Rodriguez} L.~F.,  {Marti} J.,
  {Robinson} C.~R.,  {Swank} J.,   {Geballe} T.,  1998, \aap, \href
  {http://adsabs.harvard.edu/cgi-bin/nph-bib_query?bibcode=1998A%26A...330L...9M&db_key=AST}
  {330, L9}

\bibitem[\protect\citeauthoryear{{Mirabel}, {Dhawan}, {Mignani}, {Rodrigues}
  \& {Guglielmetti}}{{Mirabel} et~al.}{2001}]{Mirabeletal2001}
{Mirabel} I.~F.,  {Dhawan} V.,  {Mignani} R.~P.,  {Rodrigues} I.,
  {Guglielmetti} F.,  2001, \mn@doi [\nat] {10.1038/35093060}, \href
  {https://ui.adsabs.harvard.edu/abs/2001Natur.413..139M} {413, 139}

\bibitem[\protect\citeauthoryear{{Miroshnichenko}, {Klochkova}, {Bjorkman}  \&
  {Panchuk}}{{Miroshnichenko} et~al.}{2002}]{Miroshnichenko2002}
{Miroshnichenko} A.~S.,  {Klochkova} V.~G.,  {Bjorkman} K.~S.,   {Panchuk}
  V.~E.,  2002, \mn@doi [\aap] {10.1051/0004-6361:20020798}, \href
  {https://ui.adsabs.harvard.edu/abs/2002A&A...390..627M} {390, 627}

\bibitem[\protect\citeauthoryear{{Monnier} et~al.,}{{Monnier}
  et~al.}{2007}]{Monnieretal2007}
{Monnier} J.~D.,  et~al., 2007, \mn@doi [Science] {10.1126/science.1143205},
  \href {http://cdsads.u-strasbg.fr/abs/2007Sci...317..342M} {317, 342}

\bibitem[\protect\citeauthoryear{{Morris}, {Agol}, {Davenport}  \&
  {Hawley}}{{Morris} et~al.}{2018}]{Morris2018}
{Morris} B.~M.,  {Agol} E.,  {Davenport} J. R.~A.,   {Hawley} S.~L.,  2018,
  \mn@doi [\mnras] {10.1093/mnras/sty568}, \href
  {https://ui.adsabs.harvard.edu/abs/2018MNRAS.476.5408M} {476, 5408}

\bibitem[\protect\citeauthoryear{{Motta} et~al.,}{{Motta}
  et~al.}{2017}]{Mottaetal2017}
{Motta} S.~E.,  et~al., 2017, \mn@doi [\mnras] {10.1093/mnras/stx1699}, \href
  {https://ui.adsabs.harvard.edu/abs/2017MNRAS.471.1797M} {471, 1797}

\bibitem[\protect\citeauthoryear{{Mu{\~n}oz-Darias}, {Motta}  \&
  {Belloni}}{{Mu{\~n}oz-Darias} et~al.}{2011}]{MunozDariasetal2011}
{Mu{\~n}oz-Darias} T.,  {Motta} S.,   {Belloni} T.~M.,  2011, \mn@doi [\mnras]
  {10.1111/j.1365-2966.2010.17476.x}, \href
  {https://ui.adsabs.harvard.edu/abs/2011MNRAS.410..679M} {410, 679}

\bibitem[\protect\citeauthoryear{{O'Brien}, {Horne}, {Hynes}, {Chen}, {Haswell}
   \& {Still}}{{O'Brien} et~al.}{2002}]{OBrienetal2002}
{O'Brien} K.,  {Horne} K.,  {Hynes} R.~I.,  {Chen} W.,  {Haswell} C.~A.,
  {Still} M.~D.,  2002, \mn@doi [MNRAS] {10.1046/j.1365-8711.2002.05530.x},
  \href {http://adsabs.harvard.edu/abs/2002MNRAS.334..426O} {334, 426}

\bibitem[\protect\citeauthoryear{{Orosz} et~al.,}{{Orosz}
  et~al.}{2001}]{Oroszetal2001}
{Orosz} J.~A.,  et~al., 2001, ApJ, \href
  {http://adsabs.harvard.edu/cgi-bin/nph-bib_query?bibcode=2001ApJ...555..489O&db_key=AST}
  {555, 489}

\bibitem[\protect\citeauthoryear{{Orosz}, {Steiner}, {McClintock}, {Torres},
  {Remillard}, {Bailyn}  \& {Miller}}{{Orosz} et~al.}{2011a}]{Oroszetal2011a}
{Orosz} J.~A.,  {Steiner} J.~F.,  {McClintock} J.~E.,  {Torres} M. A.~P.,
  {Remillard} R.~A.,  {Bailyn} C.~D.,   {Miller} J.~M.,  2011a, \mn@doi [\apj]
  {10.1088/0004-637X/730/2/75}, \href
  {https://ui.adsabs.harvard.edu/abs/2011ApJ...730...75O} {730, 75}

\bibitem[\protect\citeauthoryear{{Orosz}, {McClintock}, {Aufdenberg},
  {Remillard}, {Reid}, {Narayan}  \& {Gou}}{{Orosz}
  et~al.}{2011b}]{Oroszetal2011b}
{Orosz} J.~A.,  {McClintock} J.~E.,  {Aufdenberg} J.~P.,  {Remillard} R.~A.,
  {Reid} M.~J.,  {Narayan} R.,   {Gou} L.,  2011b, \mn@doi [\apj]
  {10.1088/0004-637X/742/2/84}, \href
  {https://ui.adsabs.harvard.edu/abs/2011ApJ...742...84O} {742, 84}

\bibitem[\protect\citeauthoryear{{Paice} et~al.,}{{Paice}
  et~al.}{2019}]{Paiceetal2019}
{Paice} J.~A.,  et~al., 2019, \mnras, "in press"

\bibitem[\protect\citeauthoryear{{Perlman} et~al.,}{{Perlman}
  et~al.}{2011}]{Perlman2011}
{Perlman} E.~S.,  et~al., 2011, \mn@doi [\apj] {10.1088/0004-637X/743/2/119},
  \href {https://ui.adsabs.harvard.edu/abs/2011ApJ...743..119P} {743, 119}

\bibitem[\protect\citeauthoryear{{Plotkin}, {Markoff}, {Kelly}, {K{\"o}rding}
  \& {Anderson}}{{Plotkin} et~al.}{2012}]{Plotkinetal2012}
{Plotkin} R.~M.,  {Markoff} S.,  {Kelly} B.~C.,  {K{\"o}rding} E.,   {Anderson}
  S.~F.,  2012, \mn@doi [MNRAS] {10.1111/j.1365-2966.2011.19689.x}, \href
  {http://cdsads.u-strasbg.fr/abs/2012MNRAS.419..267P} {419, 267}

\bibitem[\protect\citeauthoryear{{Rahoui}, {Lee}, {Heinz}, {Hines},
  {Pottschmidt}, {Wilms}  \& {Grinberg}}{{Rahoui} et~al.}{2011}]{Rahoui2011}
{Rahoui} F.,  {Lee} J.~C.,  {Heinz} S.,  {Hines} D.~C.,  {Pottschmidt} K.,
  {Wilms} J.,   {Grinberg} V.,  2011, \mn@doi [\apj]
  {10.1088/0004-637X/736/1/63}, \href
  {https://ui.adsabs.harvard.edu/abs/2011ApJ...736...63R} {736, 63}

\bibitem[\protect\citeauthoryear{{Rees}}{{Rees}}{1978}]{Rees1978}
{Rees} M.~J.,  1978, \mn@doi [\nat] {10.1038/275516a0}, \href
  {https://ui.adsabs.harvard.edu/abs/1978Natur.275..516R} {275, 516}

\bibitem[\protect\citeauthoryear{{Rodriguez} et~al.,}{{Rodriguez}
  et~al.}{2015}]{Rodriguezetal2015b}
{Rodriguez} J.,  et~al., 2015, \mn@doi [\apj] {10.1088/0004-637X/807/1/17},
  \href {https://ui.adsabs.harvard.edu/abs/2015ApJ...807...17R} {807, 17}

\bibitem[\protect\citeauthoryear{{Romero}, {Boettcher}, {Markoff}  \&
  {Tavecchio}}{{Romero} et~al.}{2017}]{Romeroetal2017}
{Romero} G.~E.,  {Boettcher} M.,  {Markoff} S.,   {Tavecchio} F.,  2017,
  \mn@doi [\ssr] {10.1007/s11214-016-0328-2}, \href
  {http://cdsads.u-strasbg.fr/abs/2017SSRv..207....5R} {207, 5}

\bibitem[\protect\citeauthoryear{{Rushton} et~al.,}{{Rushton}
  et~al.}{2017}]{Rushtonetal2017}
{Rushton} A.~P.,  et~al., 2017, \mn@doi [\mnras] {10.1093/mnras/stx526}, \href
  {https://ui.adsabs.harvard.edu/abs/2017MNRAS.468.2788R} {468, 2788}

\bibitem[\protect\citeauthoryear{{Russell} \& {Shahbaz}}{{Russell} \&
  {Shahbaz}}{2014}]{RussellShahbaz2014}
{Russell} D.~M.,  {Shahbaz} T.,  2014, \mn@doi [\mnras]
  {10.1093/mnras/stt2330}, \href
  {https://ui.adsabs.harvard.edu/abs/2014MNRAS.438.2083R} {438, 2083}

\bibitem[\protect\citeauthoryear{{Russell}, {Fender}, {Hynes}, {Brocksopp},
  {Homan}, {Jonker}  \& {Buxton}}{{Russell} et~al.}{2006}]{Russelletal2006}
{Russell} D.~M.,  {Fender} R.~P.,  {Hynes} R.~I.,  {Brocksopp} C.,  {Homan} J.,
   {Jonker} P.~G.,   {Buxton} M.~M.,  2006, \mn@doi [MNRAS]
  {10.1111/j.1365-2966.2006.10756.x}, \href
  {http://adsabs.harvard.edu/cgi-bin/nph-bib_query?bibcode=2006MNRAS.371.1334R&db_key=AST}
  {371, 1334}

\bibitem[\protect\citeauthoryear{{Russell}, {Maitra}, {Dunn}  \&
  {Markoff}}{{Russell} et~al.}{2010}]{Russelletal2010}
{Russell} D.~M.,  {Maitra} D.,  {Dunn} R.~J.~H.,   {Markoff} S.,  2010, \mn@doi
  [MNRAS] {10.1111/j.1365-2966.2010.16547.x}, \href
  {http://cdsads.u-strasbg.fr/abs/2010MNRAS.405.1759R} {405, 1759}

\bibitem[\protect\citeauthoryear{{Russell} et~al.,}{{Russell}
  et~al.}{2013}]{Russelletal2013}
{Russell} D.~M.,  et~al., 2013, \mn@doi [MNRAS] {10.1093/mnras/sts377}, \href
  {http://cdsads.u-strasbg.fr/abs/2013MNRAS.429..815R} {429, 815}

\bibitem[\protect\citeauthoryear{{Russell}, {Soria}, {Miller-Jones}, {Curran},
  {Markoff}, {Russell}  \& {Sivakoff}}{{Russell}
  et~al.}{2014}]{Russelletal2014}
{Russell} T.~D.,  {Soria} R.,  {Miller-Jones} J.~C.~A.,  {Curran} P.~A.,
  {Markoff} S.,  {Russell} D.~M.,   {Sivakoff} G.~R.,  2014, \mn@doi [\mnras]
  {10.1093/mnras/stt2498}, \href
  {http://cdsads.u-strasbg.fr/abs/2014MNRAS.439.1390R} {439, 1390}

\bibitem[\protect\citeauthoryear{{Russell} et~al.,}{{Russell}
  et~al.}{2015}]{Russelletal2015}
{Russell} T.~D.,  et~al., 2015, \mn@doi [\mnras] {10.1093/mnras/stv723}, \href
  {http://cdsads.u-strasbg.fr/abs/2015MNRAS.450.1745R} {450, 1745}

\bibitem[\protect\citeauthoryear{{Russell} et~al.,}{{Russell}
  et~al.}{2018}]{Russell2018ATel11533}
{Russell} D.~M.,  et~al., 2018, The Astronomer's Telegram, \href
  {https://ui.adsabs.harvard.edu/abs/2018ATel11533....1R} {11533, 1}

\bibitem[\protect\citeauthoryear{{Russell}, {Casella}, {Kalemci}, {Vahdat
  Motlagh}, {Saikia}, {Pirbhoy}  \& {Maitra}}{{Russell}
  et~al.}{2020}]{Russell2020}
{Russell} D.~M.,  {Casella} P.,  {Kalemci} E.,  {Vahdat Motlagh} A.,  {Saikia}
  P.,  {Pirbhoy} S.~F.,   {Maitra} D.,  2020, arXiv e-prints, \href
  {https://ui.adsabs.harvard.edu/abs/2020arXiv200208399R} {p. arXiv:2002.08399}

\bibitem[\protect\citeauthoryear{{Saikia}, {Russell}, {Bramich},
  {Miller-Jones}, {Baglio}  \& {Degenaar}}{{Saikia} et~al.}{2019}]{Saikia2019}
{Saikia} P.,  {Russell} D.~M.,  {Bramich} D.~M.,  {Miller-Jones} J. C.~A.,
  {Baglio} M.~C.,   {Degenaar} N.,  2019, \mn@doi [\apj]
  {10.3847/1538-4357/ab4a09}, \href
  {https://ui.adsabs.harvard.edu/abs/2019ApJ...887...21S} {887, 21}

\bibitem[\protect\citeauthoryear{{Sams}, {Eckart}  \& {Sunyaev}}{{Sams}
  et~al.}{1996}]{SamsEckartSunyaev1996}
{Sams} B.~J.,  {Eckart} A.,   {Sunyaev} R.,  1996, \mn@doi [Nature]
  {10.1038/382047a0}, \href
  {http://adsabs.harvard.edu/cgi-bin/nph-bib_query?bibcode=1996Natur.382...47S&db_key=AST}
  {382, 47}

\bibitem[\protect\citeauthoryear{{Shahbaz}, {Watson}  \& {Dhillon}}{{Shahbaz}
  et~al.}{2014}]{Shahbazetal2014}
{Shahbaz} T.,  {Watson} C.~A.,   {Dhillon} V.~S.,  2014, \mn@doi [\mnras]
  {10.1093/mnras/stu267}, \href
  {https://ui.adsabs.harvard.edu/abs/2014MNRAS.440..504S} {440, 504}

\bibitem[\protect\citeauthoryear{{Stevens} et~al.,}{{Stevens}
  et~al.}{2018}]{Stevensetal2018}
{Stevens} A.~L.,  et~al., 2018, \mn@doi [\apjl] {10.3847/2041-8213/aae1a4},
  \href {https://ui.adsabs.harvard.edu/abs/2018ApJ...865L..15S} {865, L15}

\bibitem[\protect\citeauthoryear{{Stirling}, {Spencer}, {de la Force},
  {Garrett}, {Fender}  \& {Ogley}}{{Stirling} et~al.}{2001}]{Stirlingetal2001}
{Stirling} A.~M.,  {Spencer} R.~E.,  {de la Force} C.~J.,  {Garrett} M.~A.,
  {Fender} R.~P.,   {Ogley} R.~N.,  2001, MNRAS, 327, 1273

\bibitem[\protect\citeauthoryear{{Swain} et~al.,}{{Swain}
  et~al.}{2003}]{swain2003}
{Swain} M.,  et~al., 2003, \mn@doi [\apjl] {10.1086/379235}, \href
  {https://ui.adsabs.harvard.edu/abs/2003ApJ...596L.163S} {596, L163}

\bibitem[\protect\citeauthoryear{{Tetarenko} et~al.,}{{Tetarenko}
  et~al.}{2017}]{Tetarenkoetal2017}
{Tetarenko} A.~J.,  et~al., 2017, \mn@doi [\mnras] {10.1093/mnras/stx1048},
  \href {https://ui.adsabs.harvard.edu/abs/2017MNRAS.469.3141T} {469, 3141}

\bibitem[\protect\citeauthoryear{{Thureau} et~al.,}{{Thureau}
  et~al.}{2009}]{Thureauetal2009}
{Thureau} N.~D.,  et~al., 2009, \mn@doi [\mnras]
  {10.1111/j.1365-2966.2009.14949.x}, \href
  {https://ui.adsabs.harvard.edu/abs/2009MNRAS.398.1309T} {398, 1309}

\bibitem[\protect\citeauthoryear{{Tingay} et~al.,}{{Tingay}
  et~al.}{1995}]{Tingayetal1995}
{Tingay} S.~J.,  et~al., 1995, Nature, 374, 141

\bibitem[\protect\citeauthoryear{{Tomsick} \& {Muterspaugh}}{{Tomsick} \&
  {Muterspaugh}}{2010}]{TomsickMuterspaugh2010}
{Tomsick} J.~A.,  {Muterspaugh} M.~W.,  2010, \mn@doi [\apj]
  {10.1088/0004-637X/719/1/958}, \href
  {https://ui.adsabs.harvard.edu/abs/2010ApJ...719..958T} {719, 958}

\bibitem[\protect\citeauthoryear{{T{\"o}r{\"o}k}, {Bakala},
  {{\v{S}}r{\'a}mkov{\'a}}, {Stuchl{\'\i}k}  \& {Urbanec}}{{T{\"o}r{\"o}k}
  et~al.}{2010}]{Toroketal2010}
{T{\"o}r{\"o}k} G.,  {Bakala} P.,  {{\v{S}}r{\'a}mkov{\'a}} E.,
  {Stuchl{\'\i}k} Z.,   {Urbanec} M.,  2010, \mn@doi [\apj]
  {10.1088/0004-637X/714/1/748}, \href
  {https://ui.adsabs.harvard.edu/abs/2010ApJ...714..748T} {714, 748}

\bibitem[\protect\citeauthoryear{{Torres}, {Casares}, {Jim{\'e}nez-Ibarra},
  {Mu{\~n}oz-Darias}, {Armas-Padilla}, {Jonker}  \& {Heida}}{{Torres}
  et~al.}{2019}]{Torresetal2019}
{Torres} M.~A.~P.,  {Casares} J.,  {Jim{\'e}nez-Ibarra} F.,  {Mu{\~n}oz-Darias}
  T.,  {Armas-Padilla} M.,  {Jonker} P.~G.,   {Heida} M.,  2019, arXiv
  e-prints, \href {https://ui.adsabs.harvard.edu/abs/2019arXiv190700938T} {p.
  arXiv:1907.00938}

\bibitem[\protect\citeauthoryear{{Tucker} et~al.,}{{Tucker}
  et~al.}{2018}]{Tuckeretal2018}
{Tucker} M.~A.,  et~al., 2018, \mn@doi [\apjl] {10.3847/2041-8213/aae88a},
  \href {https://ui.adsabs.harvard.edu/abs/2018ApJ...867L...9T} {867, L9}

\bibitem[\protect\citeauthoryear{{Vincentelli} et~al.,}{{Vincentelli}
  et~al.}{2018}]{Vincentelli2018}
{Vincentelli} F.~M.,  et~al., 2018, \mn@doi [\mnras] {10.1093/mnras/sty710},
  \href {https://ui.adsabs.harvard.edu/abs/2018MNRAS.477.4524V} {477, 4524}

\bibitem[\protect\citeauthoryear{{Waisberg} et~al.,}{{Waisberg}
  et~al.}{2017}]{Waisberg2017}
{Waisberg} I.,  et~al., 2017, \mn@doi [\apj] {10.3847/1538-4357/aa79f1}, \href
  {https://ui.adsabs.harvard.edu/abs/2017ApJ...844...72W} {844, 72}

\bibitem[\protect\citeauthoryear{{Waisberg}, {Dexter}, {Petrucci}, {Dubus}  \&
  {Perraut}}{{Waisberg} et~al.}{2019a}]{GRAVITY2019ss433}
{Waisberg} I.,  {Dexter} J.,  {Petrucci} P.-O.,  {Dubus} G.,   {Perraut} K.,
  2019a, \mn@doi [\aap] {10.1051/0004-6361/201834746}, \href
  {https://ui.adsabs.harvard.edu/abs/2019A&A...623A..47W} {623, A47}

\bibitem[\protect\citeauthoryear{{Waisberg}, {Dexter}, {Olivier-Petrucci},
  {Dubus}  \& {Perraut}}{{Waisberg} et~al.}{2019b}]{GRAVITY2019ss433b}
{Waisberg} I.,  {Dexter} J.,  {Olivier-Petrucci} P.,  {Dubus} G.,   {Perraut}
  K.,  2019b, \mn@doi [\aap] {10.1051/0004-6361/201834747}, \href
  {https://ui.adsabs.harvard.edu/abs/2019A&A...624A.127W} {624, A127}

\bibitem[\protect\citeauthoryear{{Walker}, {Hardee}, {Davies}, {Ly}  \&
  {Junor}}{{Walker} et~al.}{2018}]{Walker2018}
{Walker} R.~C.,  {Hardee} P.~E.,  {Davies} F.~B.,  {Ly} C.,   {Junor} W.,
  2018, \mn@doi [\apj] {10.3847/1538-4357/aaafcc}, \href
  {https://ui.adsabs.harvard.edu/abs/2018ApJ...855..128W} {855, 128}

\bibitem[\protect\citeauthoryear{{Weigelt} et~al.,}{{Weigelt}
  et~al.}{2012}]{weigelt2012}
{Weigelt} G.,  et~al., 2012, \mn@doi [\aap] {10.1051/0004-6361/201219213},
  \href {https://ui.adsabs.harvard.edu/abs/2012A&A...541L...9W} {541, L9}

\bibitem[\protect\citeauthoryear{{Zanin}, {Fern{\'a}ndez-Barral}, {de O{\~n}a
  Wilhelmi}, {Aharonian}, {Blanch}, {Bosch-Ramon}  \& {Galindo}}{{Zanin}
  et~al.}{2016}]{Zaninetal2016}
{Zanin} R.,  {Fern{\'a}ndez-Barral} A.,  {de O{\~n}a Wilhelmi} E.,  {Aharonian}
  F.,  {Blanch} O.,  {Bosch-Ramon} V.,   {Galindo} D.,  2016, \mn@doi [\aap]
  {10.1051/0004-6361/201628917}, \href
  {https://ui.adsabs.harvard.edu/abs/2016A&A...596A..55Z} {596, A55}

\bibitem[\protect\citeauthoryear{{Zdziarski}, {Pjanka}, {Sikora}  \&
  {Stawarz}}{{Zdziarski} et~al.}{2014}]{Zdziarskietal2014}
{Zdziarski} A.~A.,  {Pjanka} P.,  {Sikora} M.,   {Stawarz} {\L}.,  2014,
  \mn@doi [\mnras] {10.1093/mnras/stu1009}, \href
  {https://ui.adsabs.harvard.edu/abs/2014MNRAS.442.3243Z} {442, 3243}

\bibitem[\protect\citeauthoryear{{della Valle}, {Mirabel}  \&
  {Rodriguez}}{{della Valle} et~al.}{1994}]{dellaValleetal1994}
{della Valle} M.,  {Mirabel} I.~F.,   {Rodriguez} L.~F.,  1994, \aap, \href
  {https://ui.adsabs.harvard.edu/abs/1994A&A...290..803D} {290, 803}

\bibitem[\protect\citeauthoryear{{ten Brummelaar} et~al.,}{{ten Brummelaar}
  et~al.}{2005}]{tenBrummelaar2005}
{ten Brummelaar} T.~A.,  et~al., 2005, \mn@doi [\apj] {10.1086/430729}, \href
  {https://ui.adsabs.harvard.edu/abs/2005ApJ...628..453T} {628, 453}

\bibitem[\protect\citeauthoryear{{van Belle}, {Armstrong}, {Baines}, {Llama}
  \& {Schmitt}}{{van Belle} et~al.}{2019}]{vanBelle2019}
{van Belle} G.,  {Armstrong} J.~T.,  {Baines} E.,  {Llama} J.,   {Schmitt} H.,
  2019, in \baas. p.~104

\bibitem[\protect\citeauthoryear{{van Oers} et~al.,}{{van Oers}
  et~al.}{2010}]{vanOersetal2010}
{van Oers} P.,  et~al., 2010, \mn@doi [MNRAS]
  {10.1111/j.1365-2966.2010.17339.x}, \href
  {http://cdsads.u-strasbg.fr/abs/2010MNRAS.409..763V} {409, 763}

\makeatother
\end{thebibliography}







\bsp	
\label{lastpage}
\end{document}